\documentclass[a4paper,11pt]{article}
\pdfoutput=1 

\usepackage{jheppub} 

\usepackage[T1]{fontenc} 

\usepackage{latexsym,bm,amsmath,amssymb}
\usepackage{physics}
\usepackage{color}
\usepackage{tikz}
\usepackage{bbm}
\usepackage{graphicx}
\usepackage{multirow}

\title{\boldmath Effective anisotropic dynamics in Group Field Theory cosmology}

\author[a,b]{Daniele Oriti,}
\author[a,b]{Yi-Li Wang \footnote{
		Author to whom any correspondence should be addressed.}}
\affiliation[a]{Department of Physics, Shanghai University, 99 Shangda Rd, 200444, Shanghai, P.R.China}
\affiliation[b]{Arnold Sommerfeld Center for Theoretical Physics, Ludwig-Maximilians-Universit\"at M\"unchen, Theresienstra\ss e 37, 80333, M\"unchen, Germany, EU}

\emailAdd{wang.yili@physik.uni-muenchen.de}

\abstract{We study the emergent dynamics of an anisotropic universe in the context of Group Field Theory condensate cosmology, with a scalar field playing the role of a relational clock. According to different definitions of ``isotropy'', two anisotropic condensate states are considered and the Bianchi-like dynamics of cosmological anisotropic observables, as well as their quantum fluctuations, are analysed. We find that both anisotropic states become isotropic at late time, reproducing an effective Friedmann dynamics, while anisotropies give small but non-negligible contributions at earlier times, closer to the cosmic bounce.}

\begin{document} 
\maketitle
\flushbottom
\section{Introduction}
The Standard Cosmological Model (SCM), also known as $\Lambda$CDM model, is based on classical General Relativity (in particular, its FRW solution) for the evolution of the universe as a whole, and on quantum field theory for its matter content, thus on a semi-classical framework. It also rests on the assumption that the universe is homogeneous and isotropic at large scales, i.e. the cosmological principle. The SCM is extremely successful, being supported by a wide range of observations, starting from the cosmic microwave background (CMB) from the early universe. The cosmological constant, on the other hand, provides a good fit (even if not a fundamental explanation) for the current observed accelerated expansion. Extensions of the SCM still formulated in the context of semiclassical physics, starting with inflationary models but restricted to them \cite{Brandenberger:2010dk}, account for structure formation, based on the physics of small inhomogeneities in the very early universe. \\

The SCM is also incomplete, as a fundamental theory. The initial singularity at the beginning of the universe, implied by GR \cite{Penrose,Hawking,Hawking_1970}, is a clear sign of such incompleteness. Its expected quantum resolution, requiring a theory of quantum gravity, is called for in several early universe scenarios and, more generally, quantum gravity effects would impact on initial conditions and early universe evolution in crucial ways. \\

One well-explored possibility is that, when quantum effects are considered, the space-time singularity can be resolved naturally \cite{Ashtekar_2014,DeWitt}. A non-singular quantum universe may be one in which the initial singularity is replaced by a big bounce \cite{PhysRevD.20.377,MELNIKOV1979263,NOVELLO_2008}, as found, for example, in string-inspired braneworlds \cite{NOVELLO_2008}, or in Loop Quantum Cosmology (LQC) \cite{Ashtekar_2011}, the LQG-inspired quantization of the symmetry-reduced sector of GR. Again, any such quantum cosmological scenario needs proper embedding and a suitable completion within a complete quantum gravity formalism.\\

The cosmological principle can be questioned as well, in the early universe. The Belinski–Khalatnikov–Lifshitz (BKL) conjecture \cite{VA_1971,VA_1982} suggests that a highly inhomogeneous model in which each local spatial region evolves independently, close to the cosmological singularity, and the local spatial evolution can be captured by Bianchi models \cite{bojowald_2010}, among which the Bianchi I is the simplest case with vanishing spatial curvature. A theory of quantum gravity would have to confirm this conjecture and account for the early universe dynamics of both inhomogeneities and anisotropies. 
\\

Even at large scales, SCM faces several challenges. An increasing number of observations indicates anomalies that question the exact isotropy of the large-scale universe (e.g. the dipole anomaly \cite{Secrest_2021,Secrest_2022}), even pointing to the possibility of a preferred direction \cite{Pontzen_2007,Russell_2014,zhao_2016,aluri_2022,Perivolaropoulos_2022}. 
Therefore, cosmological anisotropies need investigating also at large scales, and also from a quantum gravity perspective. \\

In fact, a quantum theory of gravity is necessary to study the very early universe \cite{Butterfield_1999,Ashtekar_2014}, but it may also be needed for a fundamental physics explanation of late universe acceleration, despite this being a large-scale effect, especially if the whole spacetime description in terms of geometry and fields turns out to be only an emergent, effective one, making separation of scales a non-fundamental principle, as suggested by a number of emergent gravity scenarios \cite{ Linnemann:2017hdo,Seiberg:2006wf,Padmanabhan:2014jta,Oriti:2013jga}. 
\\

On the other hand, cosmological observations can provide a crucial testing ground and observational guidance to candidate quantum gravity theories. A main task of all such candidate formalism is to show how the usual spacetime-based semiclassical picture combining GR and QFT on curved geometries emerges from the fundamental theory. In particular, they are called to complete the semiclassical history of the universe, explaining its origin as well as solving existing cosmological puzzles, while at the same time producing novel predictions for cosmological observations. \\

In this paper, we focus on a specific non-perturbative approach to quantum gravity, the Tensorial Group Field Theory (TGFT) formalism , and in particular on models with a rich and distinctive quantum geometric content (usually labelled simply GFTs) \cite{Oriti_2006,Krajewski_2012}. These are quantum field theories (with field domain given by a group manifold) of candidate building blocks of spacetime pictures as quantum simplices, and a generalisation of matrix models for 2d gravity. Their perturbative Feynman amplitudes take the form of lattice gravity path integrals and, in different variables, of spinfoam amplitudes. For models based on the $SU(2)$ group manifold, GFTs can be viewed also as a second quantised formulation of LQG \cite{Oriti_2013}. TGFTs, therefore, represent also a promising crossroad of several a priori independent quantum gravity formalisms. \\

In the GFT context, the effective cosmological dynamics for a continuum universe emerging from the fundamental one has been the subject of many recent developments. It can be studied already at the level of mean field hydrodynamics, for a universe described by a quantum gravity condensate state \cite{Gielen_16}. Among the many results, it has been shown that an isotropic GFT condensate state can effectively reproduce the Friedmann equation of a spatially flat FLRW model, while replacing the initial singularity with a quantum bounce \cite{Marchetti_2021, Oriti_2016}, and even producing at late times a dark-energy-like accelerated expansion (at least in a simplified setting) \cite{Oriti:2021rvm}. \\

Most analyses in GFT cosmology have focused on a homogeneous and isotropic universe. In this paper, we study the effective dynamics of anisotropic universes, generalising the underlying GFT states. We build on and extend the few previous works on anisotropic GFT cosmology \cite{Cesare_2017,Andrea_2022}, checking the evolution of observables characterizing anisotropies from the early epoch close to the bounce to the late universe dynamics, comparing it to that of a Bianchi I universe. In fact, we do so under two different characterizations of (an)isotropy, thus proving the robustness of our results.\\

The article is organised as follows. A brief introduction to Bianchi I model is given in section 2, for easier comparison of later results. In section 3, we introduce the basic elements of GFT condensates, that will be used in the analysis. In section 4 and section 5, using different definitions of ``isotropy'' in quantum states, we construct two types of anisotropic GFT condensate states, and compute quantum observables characterizing them, followed by an analysis of their effective dynamics in section 6. We conclude with a discussion of our results in section 7.

\section{Bianchi I Universe}
The Bianchi classification describes a set of spatially homogeneous models, organized in nine types characterized by their Lie algebra of Killing vectors \cite{bojowald_2010}. The simplest one is the Bianchi I model, where the spatial curvature and structure constants vanish. The diagonalised Bianchi I metric in a standard synchronous form reads
\begin{equation}
ds^2=-N(\tau)d\tau^2+a_1^2(\tau)dx^2+a_2^2(\tau)dy^2+a_3^2(\tau)dz^2,
\end{equation}
where $N(\tau)$ is a lapse function, and the scale factors satisfy $a_i(\tau)/a_j(\tau)\neq \mathrm{constant}$ thus encoding spatial anisotropy. The mean scale factor is defined by $a:=(a_1a_2a_3)^{1/3}$, and the mean Hubble parameter is
\begin{equation}
H=\frac{\dot{a}}{a}=\frac{1}{3}(H_1+H_2+H_3),
\end{equation}
where ``$\cdot$'' denotes $d/d\tau$, and
\begin{equation}
H_i=\frac{\dot{a}_i}{a_i}.
\end{equation}
The Hamiltonian constraint gives \cite{Ashtekar_2009}
\begin{equation}
H_1H_2+H_2H_3+H_3H_1=8\pi G \rho_{m},
\end{equation}
so the equation of motion for a Bianchi I universe reads
\begin{equation}\label{f1}
H^2=\frac{8\pi G}{3}\rho_{m}+\frac{\Sigma^2}{a^6},
\end{equation}
where
\begin{equation}
\Sigma^2=\frac{a^6}{18}\left[(H_1-H_2)^2+(H_2-H_3)^2+(H_3-H_1)^2\right]
\end{equation}
is the shear term encoding anisotropic degrees of freedom of geometry.\\

A different, convenient way of characterizing anisotropic aspects of geometry is to work with Misner's variables \cite{Misner}, defining the scale factors with a pair of parameters $\{\beta_+,\beta_-\}$,
\begin{eqnarray}
&&a_1(\tau)=V^{1/3}(\tau)e^{\beta_+(\tau)+\sqrt{3}\beta_-(\tau)},\\
&&a_2(\tau)=V^{1/3}(\tau)e^{\beta_+(\tau)-\sqrt{3}\beta_-(\tau)},\\
&&a_3(\tau)=V^{1/3}(\tau)e^{-2\beta_+(\tau)}.
\end{eqnarray}
One can also find that the shear term can be written as $\Sigma^2=V^2(\tau)((\beta_+')^2(\tau)+(\beta_-')^2(\tau))$.\\

In a background-independent theory, coordinate frames are not physical. Instead, one can use a matter reference frame formed by physical fields used as clocks and rods to localise an object in spacetime. Because Bianchi I is homogeneous, localization in time, thus a single clock, is enough. A simple choice for clock is a free massless scalar field $\phi$ \cite{Ashtekar_2009,Gielen_16}. The general solution for a Bianchi I universe coupled with a free massless scalar field reads \cite{Chiou_2007}
\begin{equation}
a_i(\phi)=a_{i,o}e^{\sqrt{8\pi G}\kappa_i (\phi-\phi_i)},
\end{equation}
where $\kappa_i$ and $\phi_i$ are constants.
With this choice, the relational dynamics of a Bianchi I universe with respect to a dynamical massless free scalar field reads \cite{Andrea_2022}
\begin{equation}\label{friedmann}
\left(\frac{V'(\phi)}{3V(\phi)}\right)^2=\left(\frac{d\beta_+}{d\phi}\right)^2
+\left(\frac{d\beta_-}{d\phi}\right)^2+\frac{4\pi G }{3}.
\end{equation}\\

In a Bianchi I universe, common quantities to measure the anisotropy are the shear term $\Sigma^2$ and the mean anisotropy parameter $\mathsf{A}$ \cite{jacobs1968,Bronnikov_2004,Saha2006,Saha2009}
\begin{equation}
	\mathsf{A}:=\sum_{i=1}^3\frac{H_i^2-H^2}{H^2}.
\end{equation}

At first sight, one might also use $\beta_+$ and $\beta_-$ to characterise the anisotropy, so that an isotropic spatial geometry is obtained when $\beta_+=\beta_{-}=0$. However, as long as they are constant, then space is isotropic even if $\beta_{\pm}\neq 0$. Moreover, $\{\beta_+,\beta_-\}$ do not uniquely characterise a Bianchi I space-time even in the non-constant case. For instance, one can redefine the parameters such that
\begin{eqnarray}
&\beta_+&=-\frac{1}{2}\left(\tilde{\beta}_+ +\sqrt{3}\tilde{\beta}_-\right),\\
&\beta_-&=-\frac{1}{2}\left(\sqrt{3}\tilde{\beta}_+ -\tilde{\beta}_-\right),
\end{eqnarray}
then the scale factors can be written as
\begin{eqnarray}
&&\tilde{a}_3(\tau):=a_1(\tau)=V^{\frac{1}{3}}e^{-2\tilde{\beta}_ +},\\
&&\tilde{a}_2(\tau):=a_2(\tau)=V^{\frac{1}{3}}e^{\tilde{\beta}_+ -\sqrt{3}\tilde{\beta}_-},\\
&&\tilde{a}_1(\tau):=a_3(\tau)=V^{\frac{1}{3}}e^{\tilde{\beta}_+ +\sqrt{3}\tilde{\beta}_-}.
\end{eqnarray}
The spatial metric then reads
\begin{equation}
dq^2=\tilde{a}_1^2(\tau)d\tilde{x}^2+\tilde{a}_2^2(\tau)d\tilde{y}^2+\tilde{a}_3^2(\tau)d\tilde{z}^2,
\end{equation}
where $\tilde{x}=z$, $\tilde{y}=y$ and $\tilde{z}=x$, which is equivalent with the previous one. One can verify that $[(d\beta_+/d\phi)^2+(d\beta_-/d\phi)^2]$ is a quantity that is invariant under such redefinition (while keeping the metric diagonal), and this is simply the shear term written in Misner's variables. Therefore, in this paper we will focus on the shear and the mean anisotropy parameter as the ``observables'' measuring the spatial anisotropy. The values of $\beta_+$ and $\beta_-$ will be applied as auxiliary quantities illustrating the isotropisation of a state.

\section{Quantum Gravity Condensates}
\subsection{Set-up}
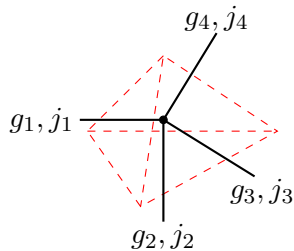
\begin{figure}[tbp]
	\centering 
	\begin{tikzpicture}
	\draw [fill] (0,0.15) circle [radius=.05];
	\draw [red,dashed](0,1)--(-1,0)--(-0.3,-1)--(1.5,0)--(0,1)--(-0.3,-1);
	\draw [red,dashed](-1,0)--(1.5,0);
	\draw [thick] (0,0.15)--(0,-1.2);
	\draw [thick] (0,0.15)--(1.2,-0.6);
	\draw [thick] (0,0.15)--(0.7,1.3);
	\draw [thick] (0,0.15)--(-1.1,0.15);
	\node at (-1.6,0.15) {$g_1,j_1$};
	\node at (0,-1.4) {$g_2,j_2$};
	\node at (1.3,-0.8) {$g_3,j_3$};
	\node at (0.7,1.5) {$g_4,j_4$};
	\end{tikzpicture}
	\caption{\label{snw} A tetrahedron (red dashed lines) corresponds to a four-valent spin network (black solid lines), where each link is dual to a face of tetrahedron.}
\end{figure}

We consider GFT models for 4d quantum gravity based on a description of quantum geometry in terms of $SU(2)$ data. The basic quanta of the field are quantum 3-simplices whose quantum geometry is encoded in four $SU(2)$ group elements (one for each of its triangular faces), and that carry also an additional real variable corresponding to a discretised (massless) quantum scalar field. This will later play the role of physical clock. The basic GFT field is thus
\begin{equation}
\varphi(g_v,\phi): SU(2)^4\times \mathbb{R}\to\mathbb{C}.
\end{equation}
The quanta of the field can also be expressed as a four-valent spin network vertex, upon Peter-Weyl expansion in $SU(2)$ representations, where each link represents one face (a triangle) of the tetrahedron, as in Fig.\ref{snw}.\\

The dynamics of these building blocks  of quantum space is governed by the partition function built from the action 
\begin{equation}
S[\varphi,\bar{\varphi}]=\int dg_v dg_w d\phi_v d\phi_w \bar{\varphi}(g_v,\phi_v)K(g_v,g_w;(\phi_v-\phi_w)^2)\varphi(g_w,\phi_w)+V,
\end{equation}
where the interaction $V$ is local in the scalar field variables, and simply identifies them across different GFT fields, while the kinetic term involves, in principle, arbitrary orders of derivatives in the scalar field variables, while being normally of at most second order in derivatives on the group manifold. 
Different specific GFT models can be considered, and the general guideline for model building is that the perturbative Feynman amplitudes of the model take the form of nice lattice gravity path integrals including the coupling with a discretised scalar field, on the lattice dual to the given Feynman diagram. Equivalently, the same amplitudes take the form of spin foam amplitudes for quantum geometry (expressed in terms of group representations) coupled to the same discretised scalar field variables.\\

In the following, we will use a general parameterised form of the dynamics, which is however fitting with the most developed spin foam models for Lorentzian quantum gravity, like the EPRL model (expressed in terms of $SU(2)$ data).\\

The GFT fields can be promoted to operators $\hat{\varphi}(g_v,\phi)$ and $\hat{\varphi}^{\dagger}(g,\phi)$, that annihilate or create quantum tetrahedra. The Fock vacuum $\ket{0}$ is defined by $\hat{\varphi}(g_v,\phi)\ket{0}=\ket{0}$, as the state with no tetrahedron, thus no geometric or topological content. Their commutator reads
\begin{equation}
[\hat{\varphi}(g_v,\phi_v),\hat{\varphi}^{\dagger}(g_w,\phi_w)]
=\delta(\phi_v-\phi_w)\int d\gamma\prod_{i=1}^{4}\delta(g_{v_i}\gamma g_{w_i}^{-1}),
\end{equation}
having chosen bosonic statistics.\\

The reconstruction of a classical continuum universe from GFTs, is based on the introduction of GFT condensate states \cite{Gielen_2014,Gielen_16},  enabling us to consider the \emph{collective} behaviour of (a very large number of) quantum tetrahedra. The simplest condensate state is characterised by a macroscopic wavefunction $\sigma(g,\phi)$ (assumed to be invariant under the diagonal left and right actions of $S(2)$ on its four group arguments) \cite{Gielen_2014,Gielen_16}
\begin{equation} \label{condensate}
\ket{\sigma}:= \mathcal{N}_{\sigma}\exp{\int dg d\phi \sigma(g,\phi)\hat{\varphi}^{\dagger}(g,\phi)}\ket{0},
\end{equation}
where $\mathcal{N}_{\sigma}=\exp{-\|\sigma\|^2/2}$, and $\|\sigma\|^2=\int dg d\phi \abs{\sigma(g,\phi)}^2=\bra{\sigma}\hat{N}\ket{\sigma}$ such that $\bra{\sigma}\ket{\sigma}=1$. It satisfies
\begin{equation}
\hat{\varphi}(g,\phi)\ket{\sigma}=\sigma(g,\phi)\ket{\sigma}.
\end{equation}
The key fact guiding (and further justifying) the cosmological interpretation of the dynamics of such states is that this wavefunction is defined over a configuration space that is isomorphic to the minisuperspace of spatially homogeneous geometries \cite{Gielen:2014ila}. \\

The effective mean field dynamics of the condensate state is encoded in the equation for the wave function obtained from the Schwinger-Dyson equation
\begin{equation}
\bra{\sigma}\frac{\delta S[\hat{\varphi},\hat{\varphi}^{\dagger}]}{\delta \hat{\varphi}^{\dagger}}\ket{\sigma}=0 .
\end{equation}
It is given by
\begin{equation}
\int dg_v d\phi_v K(g_v,g_w;(\phi_v-\phi_w)^2)\sigma(g_v,\phi_v)+\frac{\delta V}{\delta \bar{\sigma}(g_w,\phi_w)}=0.
\end{equation}
For computational simplicity in the following sections, we will work in spin representation. The wave-function decomposes as \cite{Ilkka_2019}
\begin{equation}
\sigma(g_v,\phi_v)=\sum_{\{j\},\{m\},l_L,l_R}\iota^{j_1j_2j_3j_4l_L}_{m_1m_2m_3m_4}\iota^{j_1j_2j_3j_4l_R}_{n_1n_2n_3n_4}\sigma^{j_1j_2j_3j_4l_Ll_R}(\phi_v)\prod_{i=1}^{4}D_{m_in_i}^{j_i}(g_{v,i}),
\end{equation}
where one has the intertwiner
\begin{equation}
\iota_{m_1m_2m_3m_4}^{j_1j_2j_3j_4 l}=\sum_{m,m'}C_{m_1m_2m}^{j_1j_2l}C_{m_3m_4m'}^{j_3j_4l'}C_{mm'0}^{ll'0} ,
\end{equation}
 and $C^{jj_1j_2}_{mm_1m_2}$ is the 3j-symbol. The labels $L$ and $R$ of the intertwiner $\iota$ denote the invariance under local frame rotations (left invariance) and the closure condition to make four faces (triangles) form a tetrahedron (right gauge invariance).\\

To fix the shape of a tetrahedron with given areas of its faces, we require that the intertwiner is an eigen-vector of the volume operator (obtained from quantising simplicial geometry or from LQG methods) that yields the largest possible eigenvalue, such that
\begin{equation}
    \iota^{j_1j_2j_3j_4l_R}_{n_1n_2n_3n_4}=\iota^{j_1j_2j_3j_4l_\star}_{n_1n_2n_3n_4},
\end{equation}
where $\iota^{j_1j_2j_3j_4l_\star}_{n_1n_2n_3n_4}$ is the intertwiner yielding the largest volume eigenvalue.
Then one chooses $l_L=l_R=l_\star$, and the wave-function satisfies (see \cite{Oriti_2016} for details)
\begin{equation}
\sigma(g_v,\phi_v)=\sum_{{j},m}\iota^{j_1j_2j_3j_4l_\star}_{m_1m_2m_3m_4}\iota^{j_1j_2j_3j_4l_\star}_{n_1n_2n_3n_4}\sigma^{j_1j_2j_3j_4}(\phi_v)\prod_{i=1}^{4}D_{m_in_i}^{j_i}(g_{v,i}).
\end{equation}
\\

A special choice of dynamics, used also in some previous works on GFT cosmology \cite{Cesare_2017,Pithis_2017}, is given by the kinetic kernel $K$ to be 
\begin{equation}
K=\delta(g_wg_v^{-1})\delta(\phi_v-\phi_w)\left[-\tau\partial_{\phi}^2+\eta\sum_{j_i}j_i(j_i+1)+M^2\right],
\end{equation}
and $j_i$ is the spin assigned to the $i^{\text{th}}$ link of a spin network with vertex $v$. This is consistent with a parameterised version of the GFT model \cite{Oriti:2007vf,Oriti:2010hg,Cesare_2017,Pithis_2017} based on Engle-Pereira-Rovelli-Livine (EPRL) SF model \cite{Engle:2007wy}. This choice is also consistent with the requirement of GFT renormalisability \cite{BenGeloun:2011jnm,BenGeloun:2011rc,Carrozza:2012uv,Carrozza:2013wda,Carrozza:2014rya}.


Assuming that the interaction terms give subdominant contributions to the effective dynamics (which is required by the relevance of the spin foam or lattice gravity path integral expansion of the GFT dynamics), the equation of motion becomes
\begin{equation}\label{spineqn}
A_{j_v}\partial_{\phi}^2\sigma_{j_v}-B_{j_v}\sigma_{j_v}\simeq 0,
\end{equation}
where $A_{j_v}=-\tau$, and $B_{j_v}=-(\eta\sum_{j_i}j_i(j_i+1)+M^2)$. \\


The same kind of second-order equation in the scalar field variable can be obtained, in fact, from a more general kinetic operator, for a special class of condensate states, called \lq coherent peaked states\rq.

\subsection{Coherent peaked states}

These have been introduced \cite{Marchetti_2021} in order to extract a relational dynamics in which the scalar field variable $\phi$ plays the role of clock, with respect to which the temporal evolution of geometric observables can be defined. The key idea is to choose condensate wavefunctions which are peaked on a specific value of the scalar field, which become the relevant \lq time\rq . For such condensates, higher order derivatives in the scalar field become subdominant, and the dynamics is well captured by a second order equation. A further benefit is that geometric observables constructed as expectation values (in the condensate state) of functionals of the GFT fields, and thus involving in general an integral over all values of the scalar field variables, become approximately localized on the peak value of the same, thus at a definite (if approximate) \lq time\rq . Finally, quantum fluctuations of the same relationally localized observables become well-defined, at least initially. We refer to the cited literature for more details.\\

Such coherent peaked states are given by a wavefunction
\begin{equation}
\sigma_{\epsilon}(g,\phi)=\eta_{\epsilon}(g;\phi-\phi_0,\pi_0)\tilde{\sigma}(g,\phi),
\end{equation}
where $\eta_{\epsilon}$ is a peaking function and $\tilde{\sigma}$ is the reduced wavefunction, for example:
\begin{equation}
\eta_{\epsilon}(\phi-\phi_0,\pi_0)\equiv \mathcal{N}_\epsilon\exp(-\frac{(\phi-\phi_0)^2}{2\epsilon})\exp(i\pi_0(\phi-\phi_0)),
\end{equation}
with $\mathcal{N}_{\epsilon}^2=(\pi_0\epsilon)^{-1/2}$ such that $\int d\phi \abs{\eta_{\epsilon}}^2=1$. The presence of the  $\phi_0$ term implies that quantum fluctuations do not diverge if $0 < \epsilon\ll 1$  and $\epsilon \pi_0^2\gg 1$. \\

The resulting equation of motion from the Schwinger-Dyson equation becomes (approximately) \cite{Marchetti_2021}
\begin{equation}\label{sde}
\tilde{\sigma}_{j_v}''(\phi_0)-2i\frac{\pi_0}{\epsilon \pi_0^2-1}\tilde{\sigma}_{j_v}'(\phi_0)
-\left(\epsilon^{-1}\frac{2}{\epsilon\pi_0^2-1}+\frac{B_{j_v}}{A_{j_v}}\right)\tilde{\sigma}_{j_v}(\phi_0)\simeq 0.
\end{equation}
Going to standard hydrodynamic variables $\tilde{\sigma}_{j_v}=\rho_{j_v}\exp(i\theta_{j_v})$, the equation above then yields
\begin{equation}\label{eom}
\rho_{j_v}''(\phi_0)-\frac{Q_{j_v}^2}{\rho_{j_v}^3(\phi_0)}-\Upsilon^2_{j_v}\rho_{j_v}^2(\phi_0)=0,
\end{equation}
where 
\begin{equation}
\Upsilon_{j_v}^2=\frac{\pi_0^2}{\epsilon\pi_0^2-1}\left(\frac{2}{\epsilon\pi_0^2}-\frac{1}{\epsilon\pi_0^2-1}\right)+\frac{B_{j_v}}{A_{j_v}}.
\end{equation}
Equation (\ref{eom}) has a general solution \cite{de_Cesare_2017}
\begin{equation}\label{soln}
\rho_{j_v}(\phi)=\frac{e^{\sqrt{\Upsilon_{j_v}^2} (\Phi -\phi )} \sqrt{\Omega _{j_v} e^{4 \sqrt{\Upsilon_{j_v}^2} (\phi -\Phi )}+\Omega _{j_v}-2 E_{j_v} \sqrt{\Omega_{j_v}} e^{2 \sqrt{\Upsilon_{j_v}^2} (\phi -\Phi )}}}{2 \sqrt{\Upsilon_{j_v}^2} \sqrt[4]{\Omega _{j_v}}},
\end{equation}
where $\Phi$ is the clock time read from $\phi$ when a quantum bounce happens, and
\begin{equation}
\Omega_{j_v}=E_{j_v}^2+4Q_{j_v}^2\Upsilon_{j_v}^2.
\end{equation}
There are three conserved quantities due to the introduction of peaking states \cite{Marchetti_2021}. The first is charge $Q_j$ related with $U(1)$ symmetry, which reads
\begin{equation}
Q_{j_v}=\rho_{j_v}^2\left(\theta_j'-\frac{\pi_0}{\epsilon\pi_0^2-1}\right).
\end{equation}
The others are ``bulk GFT energy'' $E_{j_v}$, a charge generating translations of the reduced background wavefunction along the clock time direction $\phi_0$, which takes the form
\begin{equation}
E_{j_v}=(\rho_{j_v}')^2+\frac{Q_{j_v}^2}{\rho_j^2}-\Upsilon_{j_v}^2\rho_{j_v}^2,
\end{equation}
and ``relational energy'' $\mathfrak{E}_{j_v}$
\begin{equation}
\mathfrak{E}_{j_v}=E_{j_v}+2Q_{j_v}\frac{\pi_0}{\epsilon\pi_0^2-1},
\end{equation}
where an ``energy injection'' comes from the precise choice of the peaking function $\eta_{\epsilon}$. 

\subsection{Observables}
We expect that the GFT condensates can reproduce classical cosmological dynamics effectively in continuum limit. To compare with the dynamics predicted by GR, and more generally to extract physical consequences from the abstract GFT dynamics, we need to compute geometrically meaningful observables and their evolution. A generic GFT $(1+1)$-body operator reads (assuming normal ordering)
\begin{equation}
\hat{O}=\int dg d\phi \hat{\varphi}^{\dagger}(g_v,\phi_v)O(g_v,g_w;\phi_v,\phi_w)\hat{\varphi}(g_w,\phi_w).
\end{equation}
The calculation of expectation values for GFT operators will be done in spin representation, which simplifies most formal manipulations, since many geometric operators are diagolanised in such basis. \\

First of all, the number operator in group representation is
\begin{equation}
\hat{N}=\int dg d\phi \hat{\varphi}^{\dagger}(g,\phi)\hat{\varphi},
\end{equation}
and its expectation value at $\phi_0$ in spin representation reads
\begin{equation}
\langle\hat{N}(\phi_0)\rangle=\int dg d\phi \abs{\sigma_{\epsilon}(g,\phi;\phi_0,\pi_0)}^2\simeq \sum_{j_v}\rho_{j_v}^2(\phi_0).
\end{equation}
One can also measure the area of surfaces associated with specific direction $i$, the macroscopic, collective counterpart of triangle areas of the individual quanta (tetrahedra):
\begin{equation}
\hat{A}_{i}=\kappa\int dgd\phi \hat{\varphi}^{\dagger}(g,\phi)\sqrt{\mathcal{E}_i^I\mathcal{E}_i^J\delta_{IJ}}\rhd\hat{\varphi}(g,\phi),
\end{equation}
where $\kappa=8\pi\gamma\ell_P$ and $\gamma$ is the Barbero-Immirzi parameter, and $\mathcal{E}_{a}^I$ is a  flux (discretised triad or normal vector to the surface) operator with $I=\{1,2,3\}$, with action 
\begin{equation}
\mathcal{E}_a^I\rhd f(g)=\lim_{\varepsilon\to 0}f(e^{-i\varepsilon\tau^I}g_1,...,g_4)
\end{equation}
for a function $f:SU(2)\to\mathbb{C}$, where $\tau^I$ is the $SU(2)$ Casimir.\\

Similarly, the volume operator reads
\begin{equation}
\hat{V}=\int dg d\phi \hat{\varphi}^{\dagger}(g_v,\phi_v)V(g_v,g_w)\hat{\varphi}^{\dagger}(g_w,\phi_w).
\end{equation}
Its expectation value is 
\begin{equation}
\langle\hat{V}(\phi_0)\rangle\simeq \sum_{j_v} V_{j_v}\rho_{j_v}^2(\phi_0),
\end{equation}
with $V_{j_v}$ the eigenvalue of volume operator (in 1st quantization) for an individual tetrahedron.\\

Finally is the scalar field operator, which can be defined as \cite{Oriti_2016}
\begin{equation}
\hat{\Phi}=\int dg_vd\phi \hat{\varphi}^{\dagger}(g_v,\phi)\hat{\varphi}(g_v,\phi)\phi.
\end{equation}
However, $\langle\hat{\Phi}\rangle$ will show a dependence on the scale of the system, i.e. $\langle\hat{\Phi}\rangle$ is extensive. The scalar field should be an intensive quantity, so a better definition is:
\begin{equation}
\hat{\phi}:=\frac{\hat{\Phi}}{\langle\hat{N}\rangle},
\end{equation}
so that
\begin{equation}
\langle\hat{\phi}\rangle=\frac{\langle\hat{\Phi}\rangle}{\langle\hat{N}\rangle}\simeq \phi_0,
\end{equation}
as expected.\\

In the same way, one can define and compute (the expectation value of) the momentum operator conjugate to the scalar field. This, together with the volume operator, allow to define the energy density for the same scalar field, of direct relevance for cosmological dynamics.

\section{Anisotropic GFT Cosmology I}

We can now proceed to study the cosmological dynamics emergent from GFT in the anisotropic case. We will do it following two different strategies for the definition of anisotropic geometries in this GFT context, in this and in the next section.

\subsection{Measure of anisotropy}
As mentioned above, we will use the shear ($[(d\beta_+/d\phi)^2+(d\beta_-/d\phi)^2]$) and anisotropy parameter $\mathsf{A}$ to measure anisotropy. However, we do not have their corresponding quantum operators in the GFT (or simplicial quantum gravity, or LQG) formalism. Therefore, we adopt the strategy of defining them as functions of areas (or volumes), which can be instead easily defined in the full theory, as we have seen, following \cite{Andrea_2022}, and then work with expressions corresponding to the classical ones, but in terms of expectation values of area/volume operators. To this end, first we will find their relations in a classical Bianchi I universe, and then we will apply them to coherent (peaked) quantum states to find their approximate quantum expectation value \cite{Andrea_2022}.\\

Before considering specific observables, though, it is necessary to define ``isotropy'' and thus ``anisotropy'' in our context, i.e. in a formalism whose only geometric information can be seen at the discrete, simplicial level.
In other words, we can only define what we mean by (an)isotropic universes, when using coherent (peaked) states, in terms of what kind of tetrahedral configurations form the GFT condensate. 
It should be clear that the essential point, for calling a tetrahedral geometry isotropic, is that it is fully specified by a single variable, the one which, in particular, specifies the tetrahedral volume. But it should also be clear that this basic fact can be realized by many types of configurations.\\

Two definitions of ``isotropy'' have been used so far in GFT. The first one uses regular (equilateral) tetrahedra to represent isotropic building blocks. This is, in fact, the most common definition, applied in most GFT cosmology \cite{Oriti_2016,Gielen_16}. The other definition uses 
tri-rectangular tetrahedra, where three orthogonal edges meeting at one vertex all have the same length \cite{Pithis_2017}. This is in line with the definition in LQC \cite{Ashtekar_2009}.\\

\begin{figure}[tbp]
	\centering 
	\begin{tikzpicture}
	\draw [->] (0,0)--(0,4);
	\draw [->] (0,0)--(5,0);
	\draw [->] (0,0)--(-2,-1.5);
	\draw [thick] (2,3.6)--(0,0)--(2,-1.5)--(4,0)--(2,3.6)--(2,-1.5);
	\draw [thick] (4,0)--(0,0);
	\node at (0,4.2) {$z$};
	\node at (-2,-1.2) {$x$};
	\node at (5.2,0) {$y$};
	\node at (-0.3,0.3) {A};
	\node at (2,-1.8) {B};
	\node at (4.2,0.3) {C};
	\node at (2,3.9) {D};
	\end{tikzpicture}
	\caption{\label{tetrahedron} An equilateral tetrahedron in the type I embedding}
\end{figure}
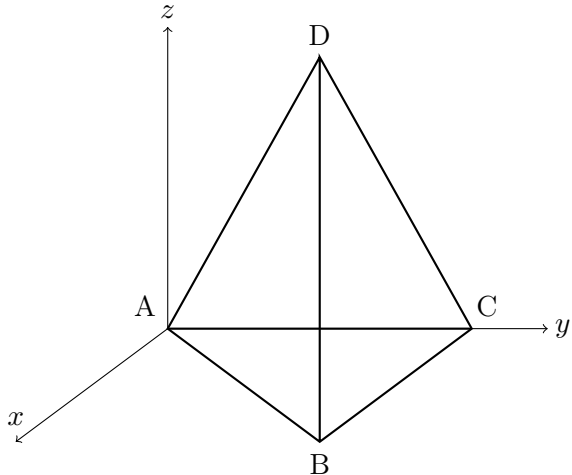

In this section, we focus on the first definition of ``isotropy''.
Suppose one has a tetrahedron with all edges of length $l$, thus also faces of equal areas, in a local embedding into flat space. 
Embedding this tetrahedron, then, in a Bianchi I spacetime will produce a rescaling of the geometric properties, in particular, its face areas, to give physical areas function of the scale factors. 
We require that one face is parallel to $x-y$ plane, where one edge of this face is also parallel to $y$-axis, so the rest $z-$axis is orthogonal to this face, as illustrated in Fig.\ref{tetrahedron}. An advantage of this embedding is that it yields a volume as a function of the areas that closest to the largest GFT eigenvalue \cite{Andrea_2022}.\\


Let the physical areas of $\bigtriangleup{ABC}$, $\bigtriangleup{ACD}$, $\bigtriangleup{ABD}$ ($\cong \bigtriangleup{BCD}$) be denoted $\mathfrak{A}$, $\mathfrak{B}$ and $\mathfrak{C}$ respectively. Their expression in terms of the Bianchi I variables read
\begin{eqnarray}
&\mathfrak{A}&=\frac{\sqrt{3}}{4}e^{2\beta_+}V^{\frac{2}{3}}l^2,\\
&\mathfrak{B}&=\frac{e^{-(\beta_++\sqrt{3}\beta_-)}}{4\sqrt{3}}\sqrt{8+e^{6\beta_++2\sqrt{3}\beta_-}}
V^{\frac{2}{3}}l^2,\\
&\mathfrak{C}&=\frac{e^{-(\beta_++\sqrt{3}\beta_-)}}{4\sqrt{3}}\sqrt{2+6 e^{4 \sqrt{3} \beta _-}+e^{6\beta_++2\sqrt{3}\beta_-}}
V^{\frac{2}{3}}l^2.
\end{eqnarray}
Then one can find $\{\beta_+,\beta_-\}$ in terms of $\mathfrak{A}$, $\mathfrak{B}$, and $\mathfrak{C}$. Because $\{\beta_+,\beta_-\}$ are ill-behaved as functions of the volume \cite{Andrea_2022}, we only consider the definition based on the areas. By requiring that $\beta_-=0$ if $\mathfrak{B}=\mathfrak{C}$ and $\beta_+=\beta_{-}=0$ if $\mathfrak{A}=\mathfrak{B}=\mathfrak{C}$, one finds
\begin{equation}\label{betam}
\beta_-=\frac{1}{4\sqrt{3}}\ln(\frac{\mathfrak{A}^2+3 \mathfrak{B}^2-12\mathfrak{C}^2}{\mathfrak{A}^2-9 \mathfrak{B}^2}),
\end{equation}
and 
\begin{equation}\label{betap}
\beta_+=\frac{1}{6}\ln(\frac{8 \mathfrak{A}^2 \sqrt{\frac{\mathfrak{A}^2+3  \mathfrak{B}^2-12 \mathfrak{C}^2}{\mathfrak{A}^2-9  \mathfrak{B}^2}}}{12\mathfrak{C}^2-\mathfrak{A}^2-3\mathfrak{B}^2}).
\end{equation}
Similarly, one can define the ratio between scale factors as functions of the areas as well,
\begin{eqnarray}
&k_1&:=\frac{a_3}{a_1}=\sqrt{\frac{1}{8}\left(9\frac{\mathfrak{B}^2}{\mathfrak{A}^2}-1\right)},\label{ratio1}\\
&k_2&:=\frac{a_3}{a_2}=\sqrt{\frac{1}{8}\left(9\frac{4\mathfrak{C}^2-\mathfrak{B}^2}{3\mathfrak{A}^2}-1\right)},\label{ratio2}
\end{eqnarray}
which are $1$ in isotropic universe. The anisotropy parameter can be expressed in two ways
\begin{eqnarray}
	&\mathsf{A}_1(\phi)&:=\frac{18 V^2(\beta_{+}'^2+\beta_{-}'^2)}{V'^2},\label{ap1}\\
	&\mathsf{A}_2(\phi)&:=\frac{2V^2(k_1^2k_2'^2+k_2^2k_1'^2-k_1k_2k_1'k_2')}{V'^2k_1^2k_2^2}.\label{ap2}
\end{eqnarray}
Finally one can find the physical volume of this tetrahedron as a function of the areas:
\begin{equation}\label{volume}
V^2=\frac{\mathfrak{A} \left(-\mathfrak{A}^2-3 \mathfrak{B}^2+12 \mathfrak{C}^2\right)}{\left(27 \sqrt{3}\right) \sqrt{\frac{\mathfrak{A}^2+3\mathfrak{B}^2-12 \mathfrak{C}^2}{\mathfrak{A}^2-9 \mathfrak{B}^2}}}.
\end{equation}

Clearly, all the relations above depend on how one embeds the tetrahedron. This ambiguity is unavoidable, but also harmless. As discussed, $\{\beta_+,\beta_-\}$ do not uniquely characterise a Bianchi I space-time. Therefore, changing embedding (for example, making $\bigtriangleup{ACD}$ parallel to $x-z$ plane) is equivalent to changing coordinate frame and thus getting a different set of $\{\beta_+,\beta_-\}$. Similar to what happens in LQC for the Bianchi I model \cite{Ashtekar_2009}, different embedding do not change the dynamics, and this is what we are really interested in.\\ 

Equipped with the information from the classical side, let us move on to consider the corresponding quantum states. The spins assigned on the links of a spin network vertex give the areas of the faces of the dual tetrahedron ($\propto\sqrt{j(j+1)}$). In the following calculations, we will substitute the quantum area eigenvalues into the classical expressions and treat the resulting expression as the eigenvalues of the corresponding quantum observables, as done also in \cite{Andrea_2022}. Moreover, areas of the triangles alone cannot specify the full shape of a tetrahedron, so we demand that each quantum building block takes the largest possible eigenvalue of its volume.\\

The eigenvalue of the volume operator is determined by the intertwiner $\iota_v$ associated to the dual spin network vertex $v$  \cite{Brunnemann_2006,Bianchi_2011,Haggard_2011,Bianchi_2012}. So strictly speaking, simply substituting eigenvalues of area operator into eqn.(\ref{volume}) does not yield the exact eigenvalue of the volume operator itself. However, the relative difference between (\ref{volume}) (obtained from the embedding according to Figure \ref{tetrahedron} ) and the exact volume eigenvalues is very small \cite{Andrea_2022}. Moreover, for our cosmological application, we only care about the expansion $V'/V$, so the qualitative result does not change if one use (\ref{volume}). 
\\

Admittedly, the complete definition of anisotropic geometric observables in quantum geometry would require use of both quantum geometric data and physical reference frame directions. In inhomogeneous case, when computing such observables in expectation values in GFT condensate state, the result would be an anisotropic observable varying with the values of the physical frame components from a relational point to another. In the homogeneous case, this dependence would be fixed once and for all, and effectively correspond to picking up a specific combination of quantum geometric data to define the anisotropic scale factors. This is then simply used without any further reference to the implicit physical frame used for their construction (unless one  studies the dependence of these observables on the evolution and backreaction of the physical frame itself, which is neglected in this paper for simplicity). 
The use of different frames would correspond to considering different such combinations of quantum geometric data. In the same approximation in which most physical aspects of the system chosen as reference frame (e.g. four scalar fields) are neglected, the same frame is expected to amount basically to a choice of embedding coordinates in a spatial manifold. In this approximation, one should also expect physical results to be invariant (at least qualitatively, with the precise quantitative aspects depending on the approximation) under different choices of coordinate frames, i.e different embeddings. 
In practice, this is what we do in this article, but without spelling out how it would result from a more complete definition of observables in the full quantum context. This more complete definition is not available in the literature, and quite non-trivial, and filling this gap in the literature is not our main purpose here. \\

\subsection{Anisotropic perturbations}
The goal is to understand now the cosmological dynamics of GFT condensates built with the above (anisotropic) building blocks. We would expect, on physical grounds, that the resulting universe has anisotropies that decrease in (clock) time. We also expect, from previous results in the GFT literature, that tetrahedral configurations with the (equal) smallest spins $j_v=1/2$ will dominate at late time \cite{Gielen_2016}. 
It is interesting to mention that a Bianchi IX space can be separated into a closed FLRW background and a gravitational wave with longest wavelength \cite{WHEELER1980,King1991}. Though we have the Bianchi I universe as our classical reference point for the construction, this is still compatible with small anisotropies, thus with a perturbative treatment \cite{Cesare_2017}. Therefore, we introduce anisotropies as ``perturbations'' . \\

Since a tetrahedron embedded as discussed above has three different faces at most, we are interested in GFT condensate wavefunctions with at most three different spins, say $(j,j,j_a,j_b)$. Modes whose $j=j_a=j_b$ form the isotropic background and the rest are the anisotropic ``perturbations''. To be specific, let $j$ corresponds to $\bigtriangleup{ABD}$ and $\bigtriangleup{BCD}$, while $j_a$ and $j_b$ gives $\bigtriangleup{ABC}$ and $\bigtriangleup{ACD}$ separately. Consequently, the order of spins is important in this paper (it was not always the case, in previous GFT literature \cite{Cesare_2017,Andrea_2022}). The order will be chosen to match with the chosen embedding.\\ 

The expectation values of GFT operators require a sum over $j_v$, which means that one has to consider all possible tetrahedra. Fortunately, this vexing task can be simplified. By convention, one usually choose $\tau<0$, $\eta>0$, and $M^2<0$. The solution (\ref{soln}) oscillates when the sum of spins on a vertex reaches some value $j_v^s$ such that $\Upsilon_{j_v^s}^2<0$, and only modes that $(j_a+j_b+2j)<j_v^s$ bring non-trivial contributions \cite{Gielen_2016}. In this section, we choose three types of background regular tetrahedra with $j_v=1/2$, $j_v=1$, and $j_v=3/2$, where $j=2$ yields a oscillating solution. Taking $\tau=-1$, one requires $-24\eta<M^2<-15\eta$. We will take $M^2=-23.9$ and $\eta=1$ to include as many modes as possible\\

We can then consider perturbations of the background modes. Using (\ref{betam}) and (\ref{betap}), one finds constraints on the perturbations, where
\begin{equation}
\frac{\mathfrak{A}^2+3 \mathfrak{B}^2-12\mathfrak{C}^2}{\mathfrak{A}^2-9 \mathfrak{B}^2}>0,
\end{equation}
which means that its quantum counter part should also satisfy
\begin{equation}
\frac{j_a(j_a+1)+3 j_b(j_b+1)-12j(j+1)}{j_a(j_a+1)-9 j_b(j_b+1)}>0.
\end{equation}
As a result, only a finite number of modes are allowed, as shown in table \ref{modes}, where we also require that the sum of spins on a spin network vertex/tetrahedron should be an integer, according to $SU(2)$ recouping theory \cite{Ilkka_2019}.\\
\begin{table}[tbp]
	\centering
	\begin{tabular}{|c|c|}
		\hline
		background&perturbations\\
		\hline 
		\multirow{2}{4em}{$(\frac{1}{2},\frac{1}{2},\frac{1}{2},\frac{1}{2})$} & $(\frac{1}{2},\frac{1}{2},\frac{1}{2},\frac{3}{2})$\\
		&$(\frac{1}{2},\frac{1}{2},1,1)$\\
		\hline
		\multirow{8}{4em}{$(1,1,1,1)$} & $(1,1,1,2)$\\
		& $(1,1,2,1)$\\
		& $(1,1,1,3)$\\
		& $(1,1,\frac{1}{2},\frac{1}{2})$\\
		& $(1,1,\frac{1}{2},\frac{3}{2})$\\
		& $(1,1,\frac{3}{2},\frac{1}{2})$\\
		& $(1,1,\frac{3}{2},\frac{3}{2})$\\
		& $(1,1,\frac{3}{2},\frac{5}{2})$\\
		\hline
		\multirow{12}{4em}{$(\frac{3}{2},\frac{3}{2},\frac{3}{2},\frac{3}{2})$}& $(\frac{3}{2},\frac{3}{2},\frac{3}{2},\frac{1}{2})$\\
		&$(\frac{3}{2},\frac{3}{2},\frac{1}{2},\frac{1}{2})$\\
		&$(\frac{3}{2},\frac{3}{2},\frac{5}{2},\frac{1}{2})$\\
		&$(\frac{3}{2},\frac{3}{2},1,1)$\\
		&$(\frac{3}{2},\frac{3}{2},2,1)$\\
		&$(\frac{3}{2},\frac{3}{2},3,1)$\\
		&$(\frac{3}{2},\frac{3}{2},\frac{1}{2},\frac{3}{2})$\\
		&$(\frac{3}{2},\frac{3}{2},\frac{5}{2},\frac{3}{2})$\\
		&$(\frac{3}{2},\frac{3}{2},1,2)$\\
		&$(\frac{3}{2},\frac{3}{2},2,2)$\\
		&$(\frac{3}{2},\frac{3}{2},\frac{3}{2},\frac{5}{2})$\\
		&$(\frac{3}{2},\frac{3}{2},1,3)$\\
		\hline
	\end{tabular}
	\caption{\label{modes} All possible modes that do not oscillate.}
\end{table}

Therefore, GFT condensates include these $25$ modes, and we assume that the wavefunction of this state is a sum over the wavefunctions governing each mode satisfying (\ref{sde}) respectively:
\begin{equation}
	\sigma=\sigma_{1/2}+\sigma_{1}+\sigma_{3/2}+\sigma_{1/2,1/2,1/2,3/2}+...+\sigma_{3/2,3/2,1,3}.
\end{equation}

Before we continue to the quantitative investigation on this model, let us do a quick analysis which will tell us the main feature that one should expect from the equation of motion.\\

Let $\tau=-1$ and $\jmath:=-\eta\sum_{i=1}^4j_i(j_i+1)-M^2$. When $\jmath<0$, solution (\ref{soln}) $\rho_{j_v}\propto \cos(\sqrt{\abs{\jmath}}\phi)$, and when $\jmath>0$, $\rho_{j_v}\propto \exp(\sqrt{\jmath}\phi)$ at large $\phi$. As a result, $\jmath$ increases as $j$ decreases, so $\abs{\sigma_{j_v}}$ grows fastest for $j=0$. Because $\abs{\sigma_{j_v}}^2=\rho_{j_v}^2$ gives the number expectation value for the mode $j_v$, $j=0$ becomes dominant for large $\abs{\phi}$. A tetrahedron with $j=0$ is usually interpreted as having a degenerate geometry, so we ignore it and set $1/2$ to be the minimum possible spin. At late time, the condensate is almost occupied by the modes whose $j=1/2$. Consequently, the wave-function
\begin{eqnarray}
\sigma(g_v,\phi)&=&\sum_{j_v}\iota_{m_1m_2m_3m_4}^{j_1j_2j_3j_4l_L}\iota_{n_1n_2n_3n_4}^{j_1j_2j_3j_4l_R}\sigma^{j_1j_2j_3j_4l_Ll_R}(\phi)\prod_{i=1}^4D_{m_in_i}^{j_i}(g_{v,i})\nonumber\\
&\simeq&\iota_{m_1m_2m_3m_4}^{1/2}\iota_{n_1n_2n_3n_4}^{1/2}\sigma_{1/2}(\phi)\prod_{i=1}^4D_{m_in_i}^{1/2}(g_{v,i})
\end{eqnarray}
for large $\phi$. Then one can approximately deal with $j=1/2$ when work in spin representation, and the kinematic kernel reads
\begin{equation}
K\simeq \delta(g_wg_v^{-1})\delta(\phi_v-\phi_w)\left[-\tau\partial_{\phi}^2+4\eta\frac{1}{2}\left(\frac{1}{2}+1\right)+M^2\right],
\end{equation}
which means that the dynamics are governed by $j=1/2$ state at late time. Therefore we expect that at large $\phi$, our anisotropic universe will become isotropic and effectively governed by a flat Friedmann equation  \cite{Gielen_16}.\\

Now let us move on to verify the statement and to see what information on the early universe can be extract from this model.

\subsection{Observables}
\begin{figure}[tbp]
	\centering 
	\includegraphics[width=0.7\textwidth]{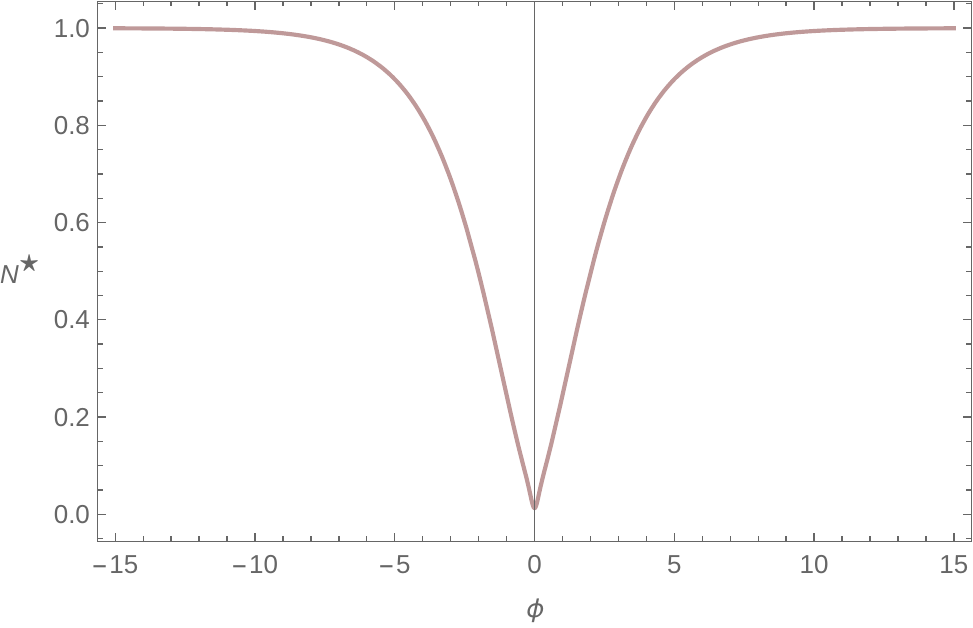}
	\caption{\label{number} The behaviour of $N_{\star}$ when $\epsilon=0.0001$ and $\pi_0=10000$. The value of $\epsilon$ brings very small slight changes to the plot.}
\end{figure}

In this part, we will show the behaviour of the number operators, of $k_i$, of $\beta_{\pm}$, and of the anisotropy parameter $\mathsf{A}$. In the following calculation, we have the energy $E_{\frac{1}{2}}=E_{1}=E_{\frac{3}{2}}=10$ with all perturbations $E_i=1$. Let 
\begin{equation}
	N_{\star}(\phi):=\frac{N_{\frac{1}{2}}(\phi)}{\langle\hat{N}(\phi)\rangle}
	=\frac{\rho_{\frac{1}{2}}(\phi)^2}{\sum_{j_v}\rho_{j_v}^2(\phi)},
\end{equation}
and it goes to $1$ at late time, which agrees with the statement that small spin dominate later, which is illustrated in Fig.\ref{number}.
\\

\begin{figure}[!h]
	\centering 
	\includegraphics[width=7.5cm]{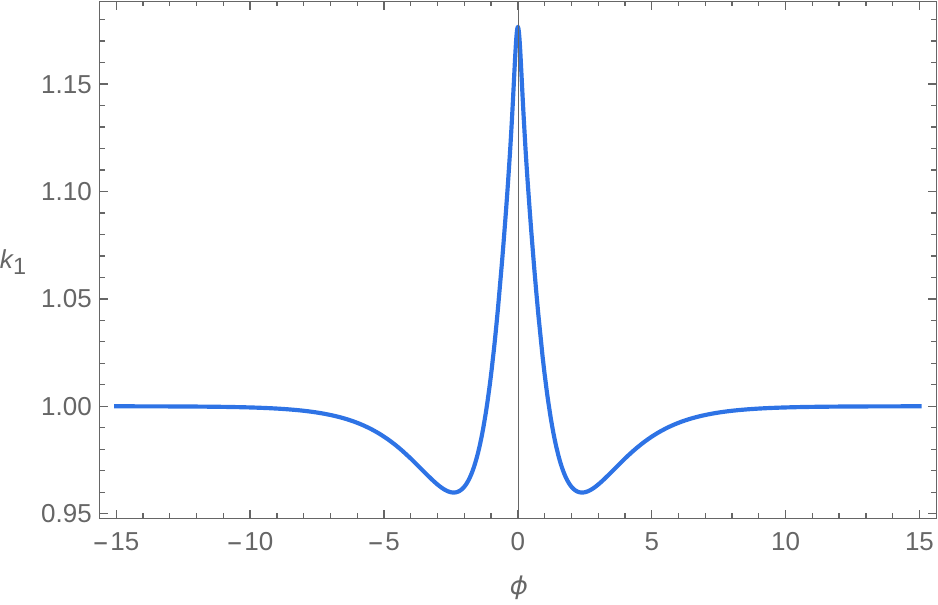}
	\includegraphics[width=7.5cm]{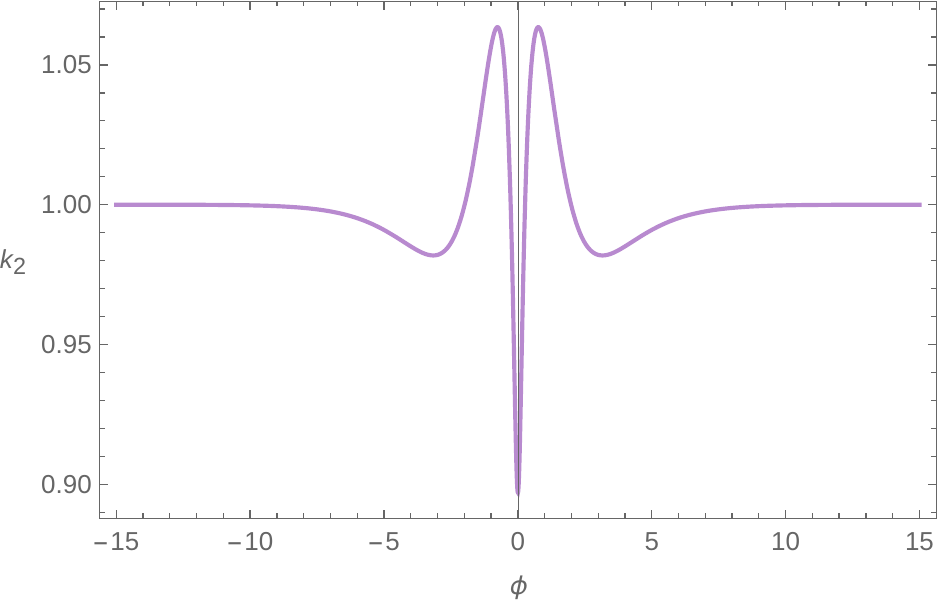}
	\caption{\label{b} The behaviour of $k_1$ and $k_2$, where $\epsilon=0.0001$ and $\pi_0=10000$.}
\end{figure}

Then for a tetrahedron with spins $j_v=(j,j,j_a,j_b)$, we express the ratio between scale factors according to (\ref{ratio1}) and (\ref{ratio2}) in terms of the spins
\begin{eqnarray}
	&&k_{1}^{j_v}=\sqrt{\frac{1}{8}\left(9\frac{j_b(j_b+1)}{j_a(j_a+1)}-1\right)},\\
	&&k_{2}^{j_v}=\sqrt{\frac{1}{8}\left(9\frac{4j(j+1)-j_b(j_b+1)}{3j_a(j_a+1)}-1\right)}.
\end{eqnarray}
Thus, we can define an observable $k_{i(\sigma)}$ to measure the anisotropy of the whole condensates, where
\begin{equation}\label{ani1}
	k_{i(\sigma)}\equiv\frac{1}{\langle\hat{N}\rangle}\sum_{j_v}(\rho_{j_v}^2k_i^{j_v}).
\end{equation}
From Fig.\ref{b}, one finds that both $k_1$ and $k_2$ go to $1$ as $\abs{\phi}$ becomes larger, which is consistent with the fact that our state is isotropic later.\\

We calculate $\{\beta_+,\beta_-\}$ in a similar way based on (\ref{betam}) and (\ref{betap}), so
\begin{eqnarray}
&&\beta_{+}^{j_v}=\frac{1}{6}\ln(\frac{8j_a(j_a+1)\sqrt{\frac{j_a(j_a+1)+3j_b(j_b+1)-12j(j+1)}{j_a(j_a+1)-9j_b(j_b+1)}}}{12j(j+1)-j_a(j_a+1)-3j_b(j_b+1)}),\\
&&\beta_{-}^{j_v}=\frac{1}{\sqrt{3}}\ln(\frac{j_a(j_a+1)+3j_b(j_b+1)-12j(j+1)}{j_a(j_a+1)-9j_b(j_b+1)}).
\end{eqnarray}

\begin{figure}[!h]
	\centering 
	\includegraphics[width=7.5cm]{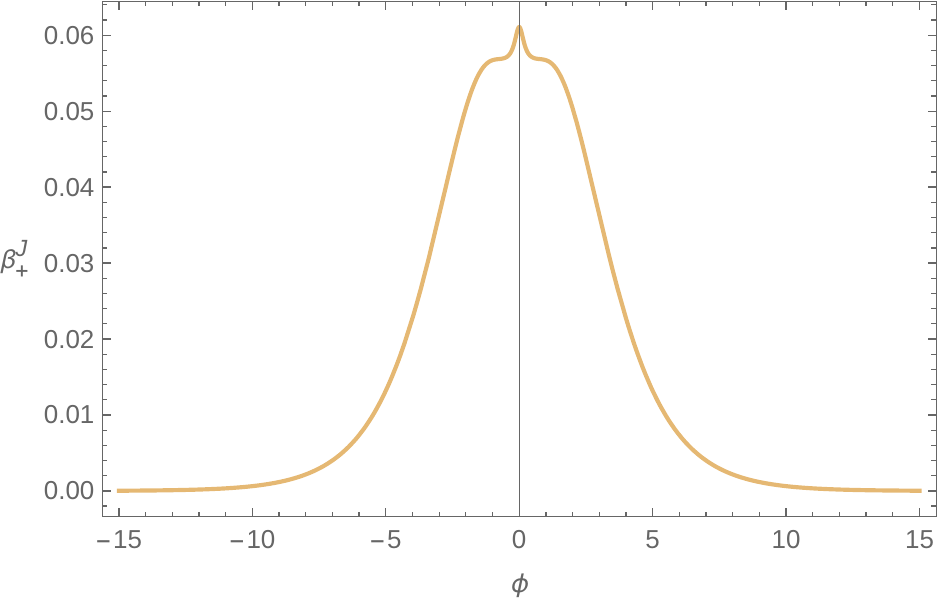}
	\includegraphics[width=7.5cm]{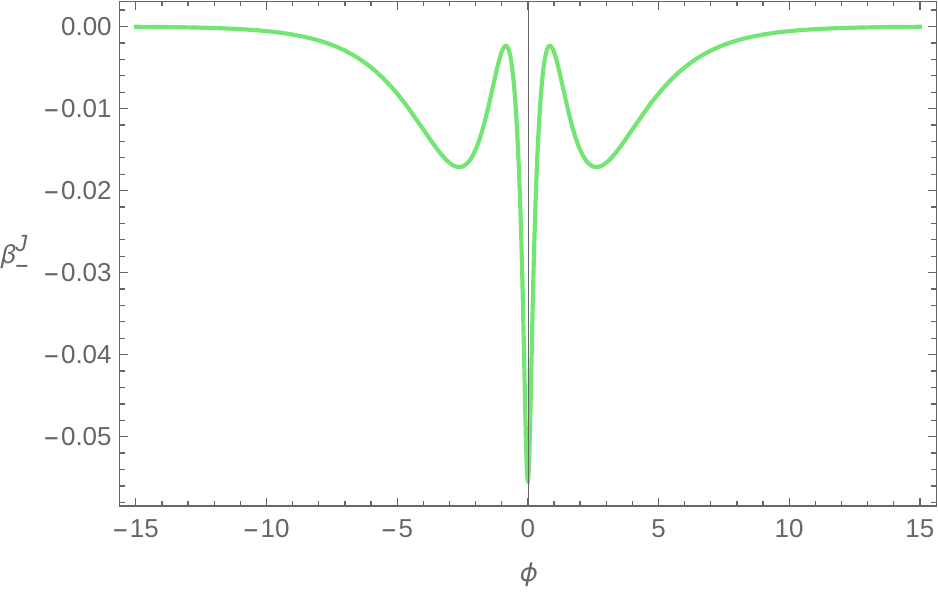}
	\caption{\label{beta} The behaviour of $\beta_{+}$ and $\beta_{-}$ in accordance with the embedding in this paper, where $\epsilon=0.0001$ and $\pi_0=10000$.}
	\end{figure}

Then let
\begin{equation}\label{ani2}
\beta_{\pm}^J\equiv\frac{1}{\langle\hat{N}\rangle}\sum_{j_v}(\rho_{j_v}^2\beta_{\pm}^{j_v}),
\end{equation}
and the result is that they become zero at late time, as is shown in Fig.\ref{beta}. \\

\begin{figure}[!h]
	\centering 
	\includegraphics[width=7.5cm]{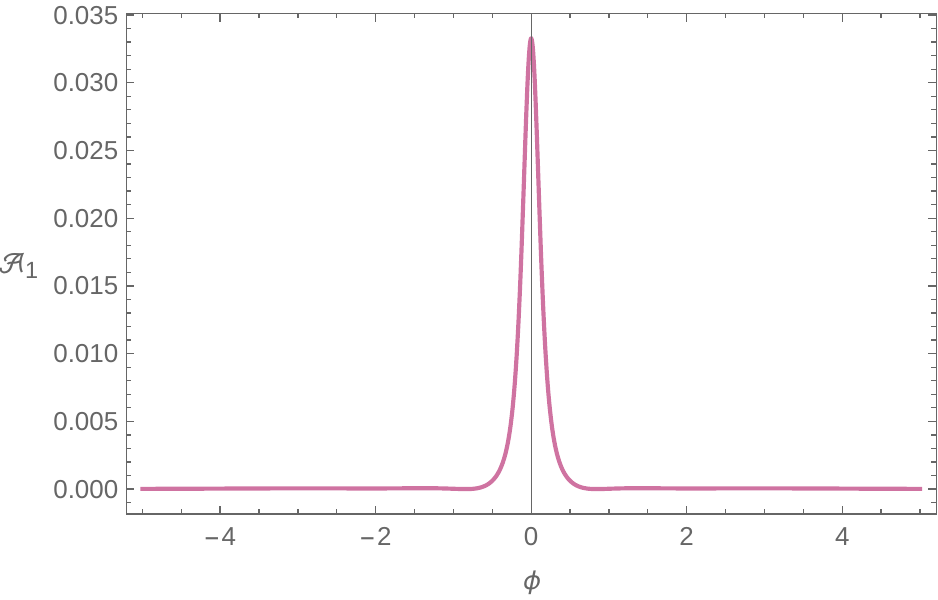}
	\includegraphics[width=7.5cm]{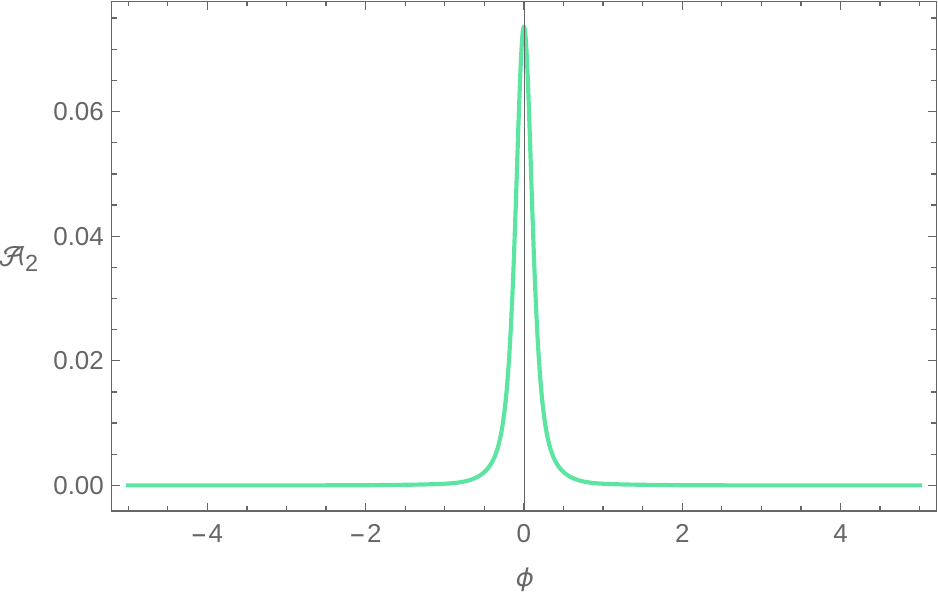}
	\caption{\label{a1} The behaviour of $\mathcal{A}_1$ and $\mathcal{A}_2$, where $\epsilon=0.0001$ and $\pi_0=10000$.}
\end{figure}

For the mean anisotropy parameter $\mathsf{A}$, one can obtain its expectation value from volume, ratio $k$, and Misners's variables according to (\ref{ap1}) and (\ref{ap2}). Let
\begin{eqnarray}
	&\mathcal{A}_1(\phi)&:=\frac{18 V^2({\beta^J_{+}}'^2+{\beta_{-}^J}'^2)}{V'^2},\\
	&\mathcal{A}_2(\phi)&:=\frac{2V^2(k_{1(\sigma)}^2k_{2(\sigma)}'^2+k_{2(\sigma)}^2k_{1(\sigma)}'^2-k_{1(\sigma)}k_{2(\sigma)}k_{1(\sigma)}'k_2')}{V'^2k_{1(\sigma)}^2k_{2(\sigma)}^2}.
\end{eqnarray}
One again finds that they goes to zero when $\abs{\phi}$ increases in Fig. \ref{a1}.\\

The calculations of these observables are consistent with each other, where all of them agree that the anisotropy is relevant near the bounce, but becomes negligible at later time.

\subsection{Quantum fluctuations}
Applying coherent peaking states, we can also copute  the relative variances of the observables, which measure the quantum fluctuations \cite{Marchetti_21}.\\

The relative variance of an observable $O$ is defined in a usual way:
\begin{equation}
		\Delta_{O}^2=\frac{\langle\hat{O}^2\rangle-\langle\hat{O}\rangle^2}{\langle\hat{O}\rangle^2}.
\end{equation}
The definition and computations are straightforward for number and volume operators. 
For the number operator, one finds \cite{Marchetti_21}
\begin{equation}
	\Delta_{N}^2=\frac{1}{\sum_{j_v}N_{j_v}}\simeq\frac{1}{\sum_{j_v}\rho_{j_v}^2}.
\end{equation}
Since $\rho_{j_v}(\phi)$ grows exponentially, the relative variance of the number operator definitely becomes negligible at later time, when the number of building blocks (on average) increases.\\

Similarly for the volume operator, one has 
\begin{equation}
\Delta_{V}^2\simeq\frac{\sum_{j_v}V_{j_v}^2\rho_{j_v}^2}{(\sum_{j_v}V_{j_v}\rho_{j_v}^2)^2},
\end{equation}
which is very small for large occupation number (which coincides with large clock time $\phi$).\\

The definition and computation of the relative variance of $k_{i}$ and $\beta_{\pm}$ are less straightforward and need more discussion. Our calculations so far apply the definition from \cite{Andrea_2022} to anisotropic measures which are obtained as functions of areas. For instance, to find $k_i$ (\ref{ani1}), one first considers every single mode and writes $k_i^{j_v}$ in terms of $j(j+1)$ (the area eigen-value of this mode). Then, one adds contributions from every mode together to get the total contribution. Such a calculation is rather similar to that of volume or area operators, which means that $k_i^{j_v}$ is also treated in a same way like an ``eigen-value'' of an ``anisotropy operator''. As a result, we expect that the relative variances of $k_{i}$ and $\beta_{\pm}$ should have 
the same behaviour, becoming negligible for large number of GFT quanta. That is,
\begin{eqnarray}
    &&\Delta_{\beta_{\pm}^J}^2\simeq\frac{1}{\langle\hat{N}\rangle^2}\frac{\sum_{j_v}(\beta^{j_v}_{\pm})^2\rho_{j_v}^2}{(\sum_{j_v}\beta^{j_v}_{\pm}\rho_{j_v}^2)^2}
    +\left(\frac{\sum_{j_v}(\rho_{j_v}^2\beta_{\pm}^{j_v})}{\langle\hat{N}\rangle^2}\right)^2\Delta_{N}^2,\\
    &&\Delta_{k_{i(\sigma)}}^2\simeq\frac{1}{\langle\hat{N}\rangle^2}\frac{\sum_{j_v}(k^{j_v}_{i})^2\rho_{j_v}^2}{(\sum_{j_v}k^{j_v}_{i}\rho_{j_v}^2)^2}+\left(\frac{\sum_{j_v}(\rho_{j_v}^2k_i^{j_v})}{\langle\hat{N}\rangle^2}\right)\Delta_{N}^2,
\end{eqnarray}
which goes to zero as the $N\to\infty$.\\

The case of $\mathcal{A}_i$ is even more complicated, as they include the derivatives with respect to the relational time $\phi$, which itself has a relative variance. If we ignore the uncertainty of these terms, and assume that there is no correlation between the volume $V$ and $\beta_{\pm}^J$ (\ref{ap1}) or $k_{i(\sigma)}$ in (\ref{ap2}) are uncorrelated. Then the relative variance of $\mathcal{A}_i$ can be found via the propagation of uncertainty. Assume that the only sources of uncertainty are $V$ and $k_{i(\sigma)}$, then
\begin{eqnarray}
    &&\Delta_{\mathcal{A}_1}^2\sim \left(\frac{V({\beta^J_{+}}'^2+{\beta_{-}^J}'^2)}{V'^2}\right)^2\Delta_V^2,\\
    &&\Delta_{\mathcal{A}_2}^2\sim \left(2\frac{\mathcal{A}_2}{V}\right)^2\Delta_V^2+\sum_ic_i^2\Delta_{k_{i(\sigma)}}^2,
\end{eqnarray}
where $c_i\equiv d\mathcal{A}_2/dk_{i(\sigma)}$ is an intensive quantity.
Hence, $\mathcal{A}_i$ also has very small relative variance at large $N$ under these assumptions.


\section{Anisotropic GFT Cosmology II}
\subsection{Measure of anisotropy}
In this section, we use the other definition of ``isotropy'' \cite{Pithis_2017}, and then compute the same anisotropy measures considered in the previous section. We consider then  tetrahedra with three edges of the same length meeting at the same vertex orthogonally, as a building block of an isotropic background.\\

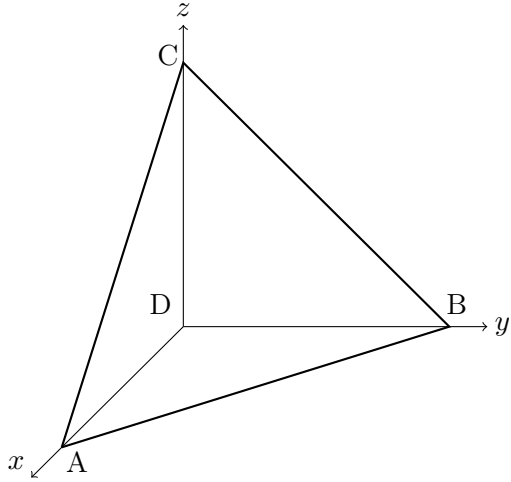
\begin{figure}[tbp]
	\centering 
	\begin{tikzpicture}
	\draw [->] (0,0)--(0,4);
	\draw [->] (0,0)--(4,0);
	\draw [->] (0,0)--(-2,-2);
	\draw [thick] (-1.6,-1.6)--(0,3.5)--(3.5,0)--(-1.6,-1.6);
	\node at (0,4.2) {$z$};
	\node at (-2.2,-1.8) {$x$};
	\node at (4.2,0) {$y$};
	\node at (-0.3,0.3) {D};
	\node at (-1.4,-1.8) {A};
	\node at (3.6,0.3) {B};
	\node at (-0.2,3.6) {C};
	\end{tikzpicture}
	\caption{\label{tetrahedron2} A tri-orthogonal tetrahedron.}
\end{figure}

Despite the different definition, the strategy to find the observables is the same, via embedding in the Bianchi I space. 
Since edges $AD$, $BD$, and $CD$ of the tetrahedron are orthogonal to each other, it is convenient to require that they are parallel to $x-$, $y-$, and $z-$ axis separately. If the length of these edges are $l$, then the physical areas of $\bigtriangleup ADB$ ($\mathfrak{A}$), $\bigtriangleup ACD$ ($\mathfrak{B}$), and $\bigtriangleup DCB$ ($\mathfrak{C}$) read
\begin{eqnarray}
	&\mathfrak{A}&=\frac{1}{2}a_1a_2l^2,\\
	&\mathfrak{B}&=\frac{1}{2}a_1a_3l^2,\\
	&\mathfrak{C}&=\frac{1}{2}a_2a_3l^2.
\end{eqnarray}
It is straightforward to find
\begin{eqnarray}
	&\mathfrak{k}_1&:=\frac{a_1}{a_3}=\frac{\mathfrak{A}}{\mathfrak{C}},\\
	&\mathfrak{k}_2&:=\frac{a_1}{a_2}=\frac{\mathfrak{B}}{\mathfrak{C}}.\\
\end{eqnarray}
Similarly, the Misner's variables $\beta_{\pm}$ can also be written in terms of the physical areas, where
\begin{eqnarray}
	&\beta_{+}&=\frac{1}{6}\ln(\frac{\mathfrak{A}^2}{\mathfrak{B}\mathfrak{C}}),\\
	&\beta_{-}&=\frac{1}{4\sqrt{3}}\ln(\frac{\mathfrak{B}^2}{\mathfrak{C}^2}).
\end{eqnarray}
Finally, the volume reads
\begin{equation}
	V=\frac{\sqrt{2}}{3}(\mathfrak{A}\mathfrak{B}\mathfrak{C})^{\frac{1}{2}}.
\end{equation}
Again, these relations only work with the chosen embedding. But as we argued already, changing embedding is equivalent to changing a coordinate frame, and should thus be immaterial in our quantum gravity context. 

\subsection{Anisotropic perturbations}
The same perturbative procedure will be repeated here, where spin networks labeled by $(1/2,1/2,1/2)$ are the isotropic background. In order to be compatible with classical description, it is required that given a spin network $(j_a,j_b,j_c)$, $j_a$, $j_b$, and $j_c$ corresponds to $\bigtriangleup ABD$ $\bigtriangleup ADC$, and $\bigtriangleup DBC$ respectively. Therefore, the order of spins is also important.\\

With this definition, one has fewer choices on the perturbations. Unlike the perturbations discussed in the previous section, here one does not have a ``background spin'' which needs to be fixed when performing perturbations, because fixing one face only corresponds to a special case where only one direction is different from the others. So $j_a$, $j_b$, and $j_c$ can be changed at the same time as a perturbation. From (\ref{soln}), one knows that the order of spins does not affect the wave-function. Meanwhile, considering all possible permutations on $(j_a,j_b,j_c)$ will result in an isotropic model, since the way we embed tetrahedra will make their contribution to the global anisotropy cancel each other. Consequently, we limit the perturbations where $j_a\leq j_b\leq j_c$ in order to obtain non-vanishing anisotropy \footnote{Strictly speaking, Bianchi I universe only show the preference at dynamical level. Here the preferred direction is picked kinetically. This weakness so far cannot be conquered since truly arbitrarily chosen spin network modes will yield a completely isotropic state. }. \\

Again, it is expected that the condensates will be dominated by small-spin modes, which is the isotropic background as well. Therefore, everything remains almost the same as in the previous section, except that here we only sum over three spins in kinetic term \cite{Pithis_2017}. Here we take $3/2$ to be maximal possible spin, such that $-M^2-45\eta/4<0$ while $-M^2-19\eta/2>0$, and ignore all oscillating solutions. The possible modes allowed in this section are shown in Table \ref{modes2}.\\

\begin{table}[tbp]
	\centering
	\begin{tabular}{|c|c|}
		\hline
		background&perturbations\\
		\hline 
		\multirow{7}{4em}{$\left(\frac{1}{2},\frac{1}{2},\frac{1}{2}\right)$} 
		& $\left(\frac{1}{2},\frac{1}{2},1\right)$\\
		& $\left(\frac{1}{2},\frac{1}{2},\frac{3}{2}\right)$\\
		& $\left(\frac{1}{2},1,1\right)$\\
		& $\left(\frac{1}{2},\frac{3}{2},\frac{3}{2}\right)$\\
		& $\left(\frac{1}{2},1,\frac{3}{2}\right)$\\
		& $\left(1,1,\frac{3}{2}\right)$\\
		& $\left(1,\frac{3}{2},\frac{3}{2}\right)$\\
		\hline
	\end{tabular}
	\caption{\label{modes2} All possible modes that do not oscillate while satisfying our conditions.}
\end{table}
 
\subsection{Observables} 
\begin{figure}[tbp]
	\centering 
	\includegraphics[width=0.6\textwidth]{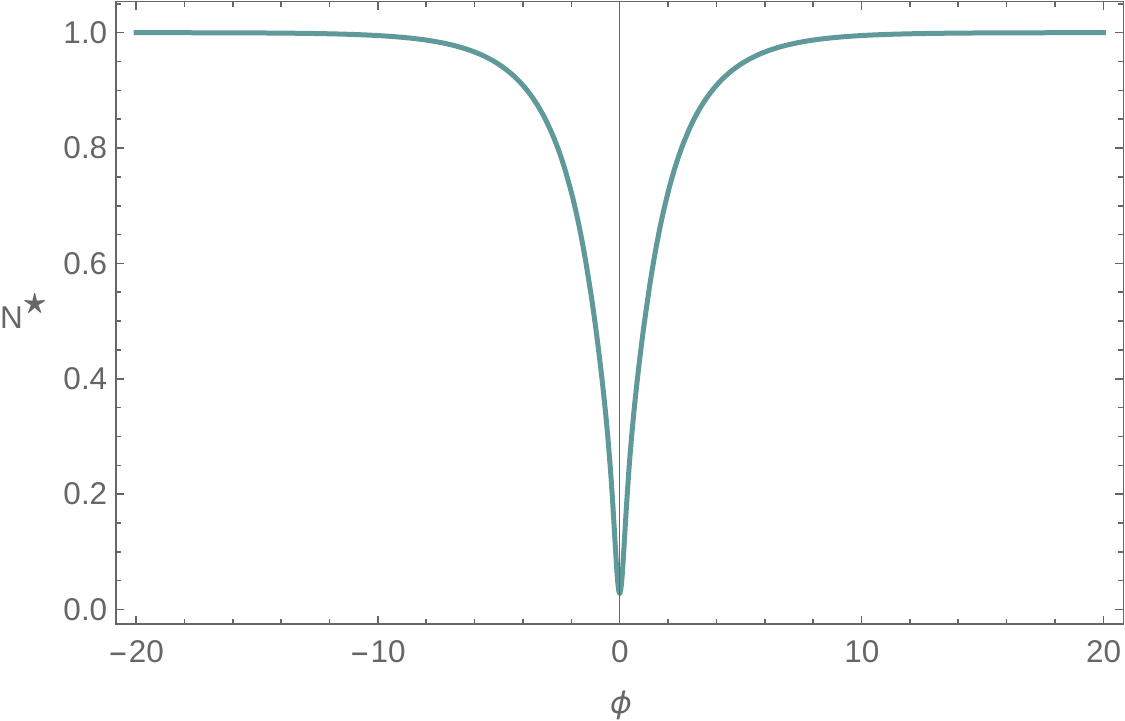}
	\caption{\label{number2} The behaviour of $N_{\star}$ when $\epsilon=0.0001$ and $\pi_0=10000$.}
\end{figure}

\begin{figure}[!h]
	\centering 
	\includegraphics[width=7.5cm]{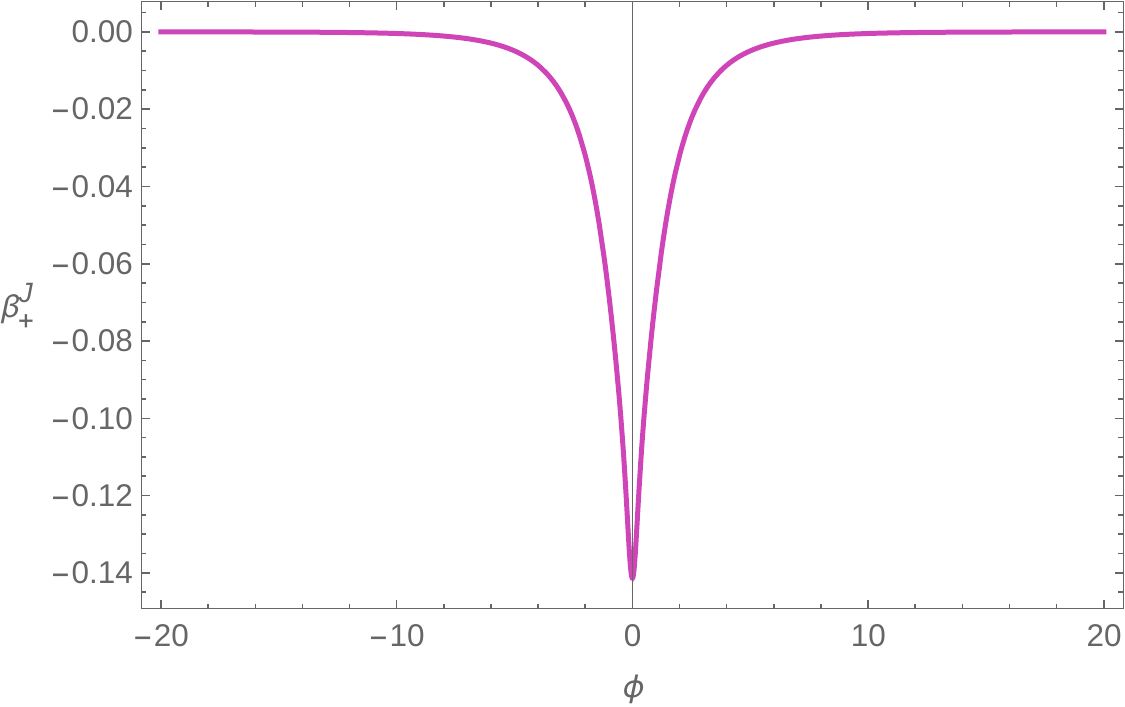}
	\includegraphics[width=7.5cm]{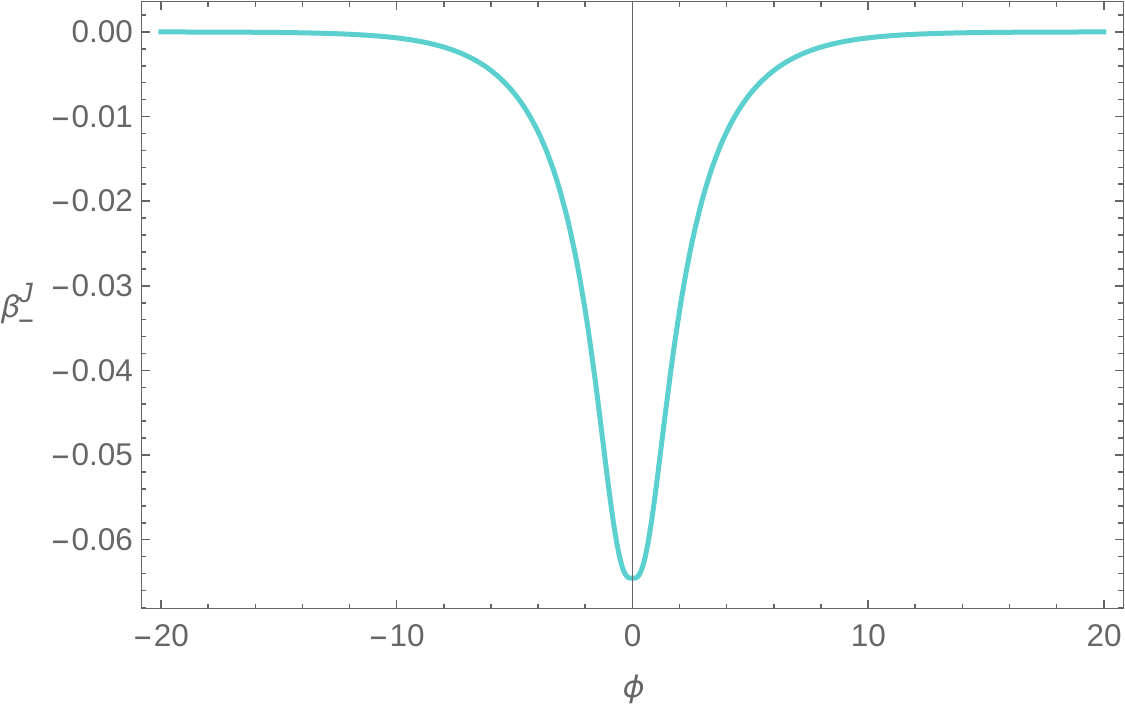}
	\caption{\label{beta2} The behaviour of $\beta_{+}$ and $\beta_{-}$ in accordance with the embedding in this paper, where $\epsilon=0.0001$ and $\pi_0=10000$.}
\end{figure}

\begin{figure}[!h]
	\centering 
	\includegraphics[width=7.5cm]{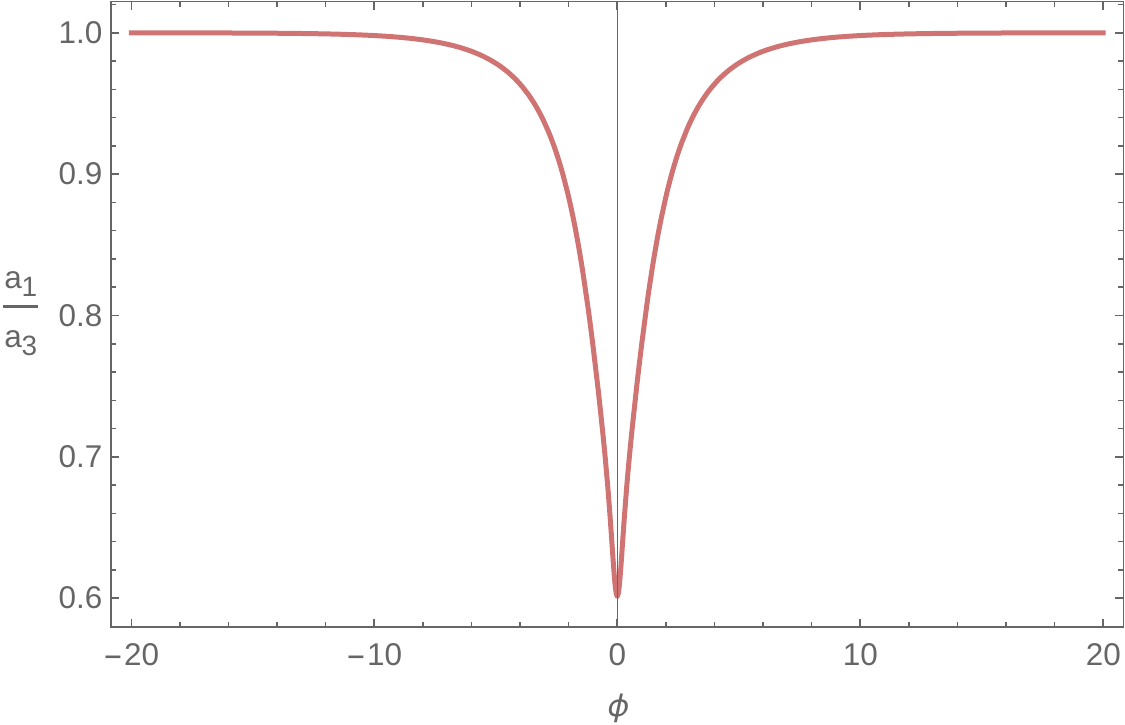}
	\includegraphics[width=7.5cm]{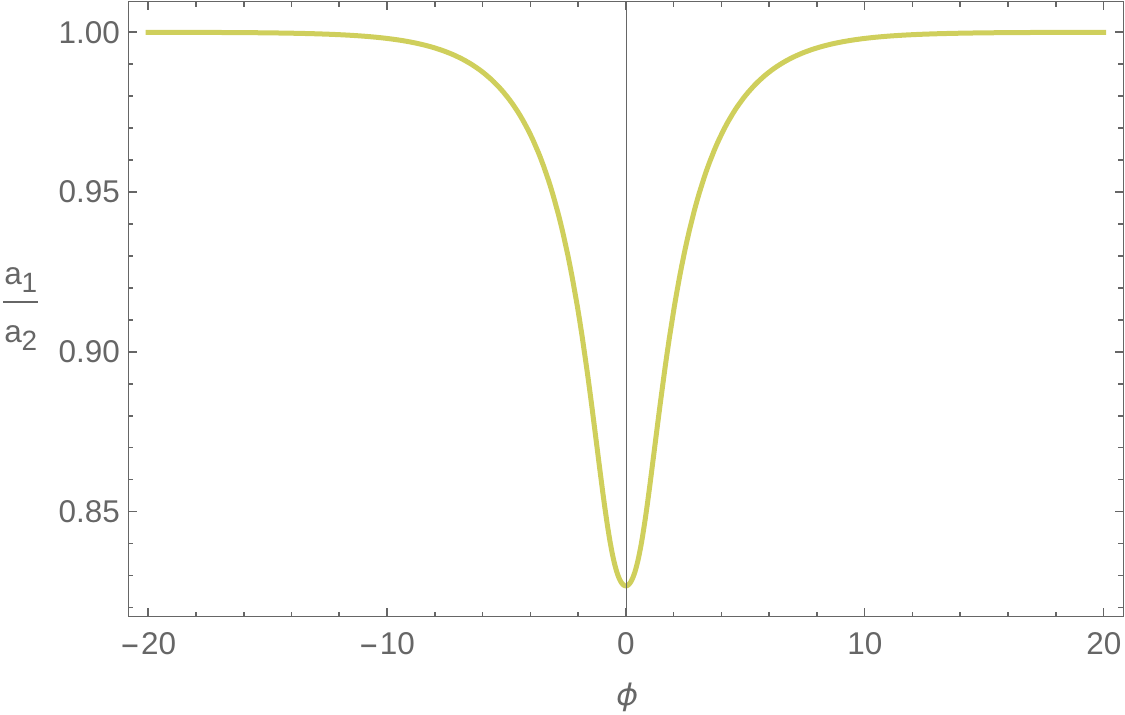}
	\caption{\label{ratio} The behaviour of $\mathfrak{k}_1$ and $\mathfrak{k}_2$ in accordance with the embedding in this paper, where $\epsilon=0.0001$ and $\pi_0=10000$.}
\end{figure}

\begin{figure}[!h]
	\centering 
	\includegraphics[width=7.5cm]{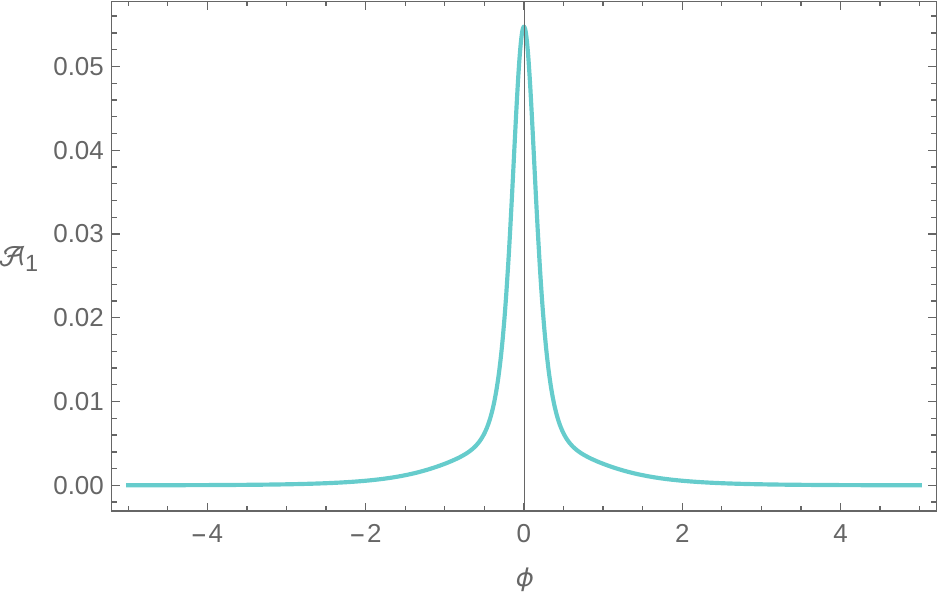}
	\includegraphics[width=7.5cm]{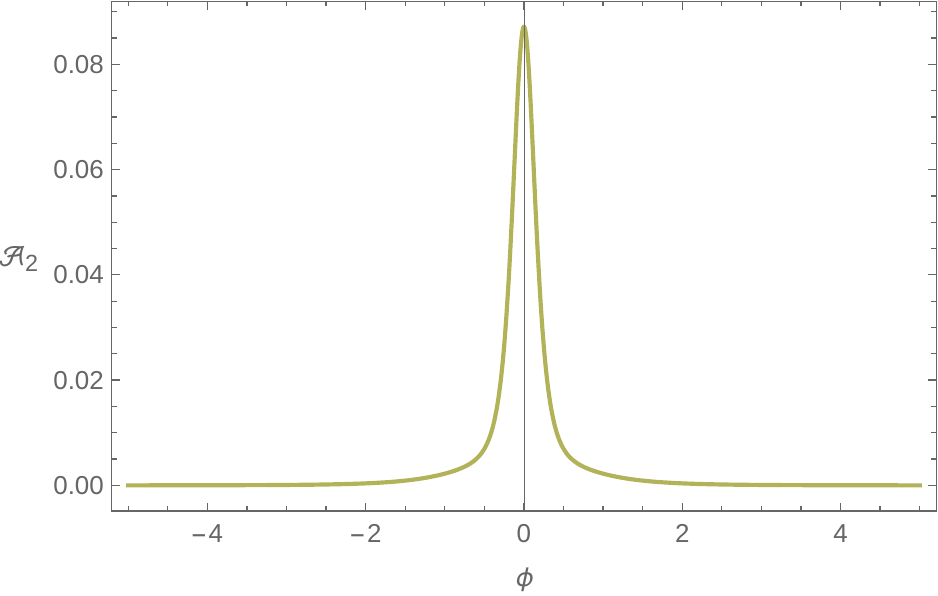}
	\caption{\label{a2} The behaviour of $\mathcal{A}_1$ and $\mathcal{A}_2$, where $\epsilon=0.0001$ and $\pi_0=10000$.}
\end{figure}

All observables are defined in the same way as in the previous section. For a tetrahedron $(j_a,j_b,j_c)$,
\begin{eqnarray}
	&\beta_{+}^{j_v}&=\frac{1}{6}\ln(\frac{j_a(j_a+1)}{\sqrt{j_b(j_b+1)j_c(j_c+1)}}),\\
	&\beta_{-}^{j_v}&=\frac{1}{4\sqrt{3}}\ln(\frac{j_b(j_b+1)}{j_c(j_c+1)}),
\end{eqnarray}
and
\begin{eqnarray}
	&\mathfrak{k}_1^{j_v}&=\frac{\sqrt{j_a(j_a+1)}}{\sqrt{j_c(j_c+1)}},\\
	&\mathfrak{k}_2^{j_v}&=\frac{\sqrt{j_b(j_b+1)}}{\sqrt{j_c(j_c+1)}}.
\end{eqnarray}
Then one can define
\begin{eqnarray}
	&&\mathfrak{k}_{i(\sigma)}=\frac{1}{\langle\hat{N}\rangle}\sum_{j_v}\rho_{j_v}^2\mathfrak{k}_i^{j_v},\\
	&&\beta_{\pm}^J=\frac{1}{\langle\hat{N}\rangle}\sum_{j_v}\rho_{j_v}^2\beta_{\pm}^{j_v}.
\end{eqnarray}
The anisotropy parameter can be expressed in the same manner as in the previous section.
Taking $-M^2=10$ and keeping all other parameters the same as in previous section, the new definition of (an)isotropy leads to the same general results as with the other definition. Again, configurations with small spins are dominant at late time, and this generically leads the condensates to become isotropic as $\phi$ grows (see Fig.\ref{number2}), \ref{beta2}, \ref{ratio}, and \ref{a2}. As the quantum perturbations have exactly the same qualitative behaviour as well, it is redundant to compute them again in this section.


\section{Effective Dynamics} 
\subsection{Generalised Friedmann equation}
Finally, let us investigate the effective anisotropic dynamics from our GFT condensates. Since
\begin{equation}
a_i(\phi)=a_{i,o}e^{\sqrt{8\pi G}\kappa_i (\phi-\phi_i)},
\end{equation}
$V'/V$ is a constant, and contribution from anisotropy on the right-hand-side of equation (\ref{friedmann}) should be a constant too. We expect that our GFT effective equation satisfy
\begin{equation}\label{eff}
\left(\frac{V'}{3V}\right)^2=\left(\frac{d\beta_{+}^J}{d\phi}\right)^2+\left(\frac{d\beta_{-}^J}{d\phi}\right)^2+\frac{4\pi G_{eff}}{3},
\end{equation}
where $G_{eff}$ is the effective gravitational constant
\begin{equation}
G_{eff}:=\frac{1}{3\pi}\left(\frac{\rho_{\frac{1}{2}}'}{\rho_{\frac{1}{2}}}\right)^2.
\end{equation}
Or equivalently, we expect
\begin{equation}\label{eff2}
\left(\frac{V'}{3V}\right)^2=\frac{\xi^2}{18}+\frac{4\pi G_{eff}}{3},
\end{equation}
where
\begin{equation}
	\xi^2=\frac{2k_{1(\sigma)}'^2}{k_{1(\sigma)}^2}+\frac{2k_{2(\sigma)}'^2}{k_{2(\sigma)}^2}
	-\frac{2k_{1(\sigma)}'k_{2(\sigma)}'}{k_{1(\sigma)}k_{2(\sigma)}},
\end{equation}
and the expression of $\xi^2$ is the same for $\mathfrak{k}_i$.\\

Note that in both isotropic and Bianchi I space-time,  $V'/V$ is a constant. So the fact that $V'/V=\mathrm{const}$ is not sufficient to figure out the precise cosmological dynamics that our GFT states yield. One has to compare two terms on the right-hand-side of eqn.(\ref{eff}) or (\ref{eff2}) in order to find out if the introduction of anisotropy affects the effective dynamics.\\

\begin{figure}[tbp]
	\centering 
	\includegraphics[width=0.5\textwidth]{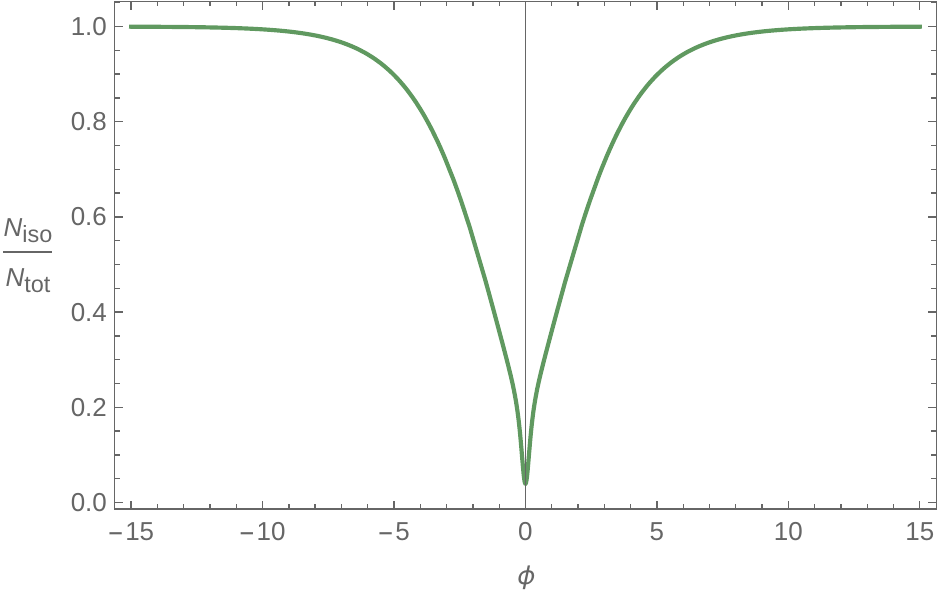}
	\caption{\label{ns} The ratio of number of isotropic tetrahedra ($N_{iso}$) over that of total tetrahedra ($N_{tot}$), where $\epsilon=0.0001$ and $\pi_0=10000$, and isotropy is represented by equilateral tetrahedra.}
\end{figure}

From Fig.\ref{ns}, one finds the isotropic (regular) background ($j_v=1/2,1,3/2$) dominates after $\phi\simeq 10$. The situation is similar for tri-rectangular condensates, where the universe becomes isotropic for $\phi\gtrsim 10$. In isotropic regime, $(d\beta_{+}/d\phi)^2+(d\beta_{-}/d\phi)^2$ or $\xi^2/18$ is negligible and 
\begin{equation}
\left(\frac{V'}{3V}\right)^2\simeq \frac{4\pi G_{eff}}{3}.
\end{equation}

\begin{figure}[!h]
	\centering 
	\includegraphics[width=11cm]{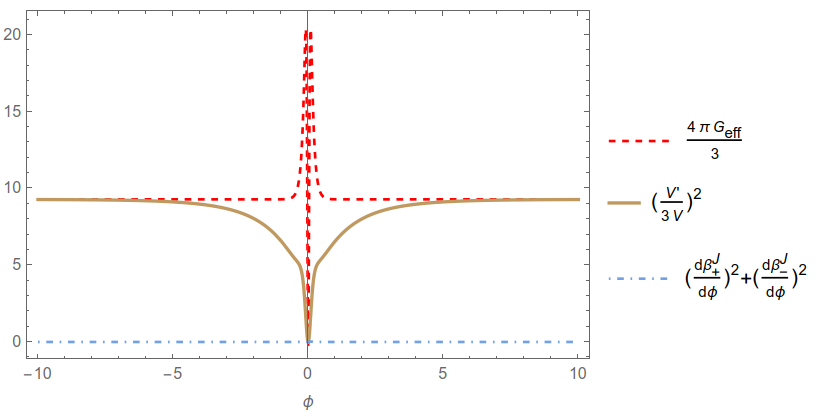}
	\includegraphics[width=10cm]{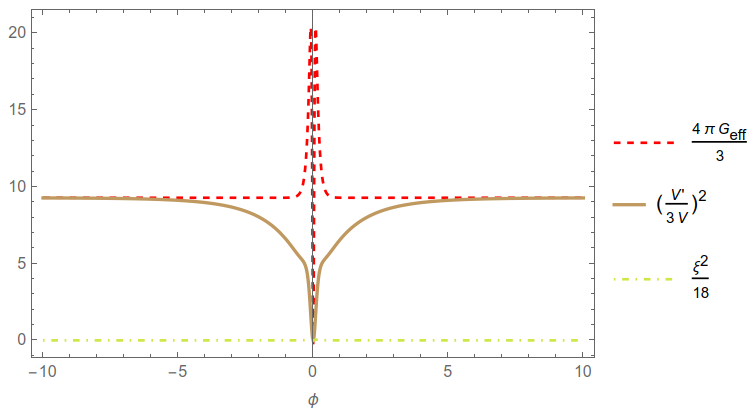}
	\caption{\label{dyn} Effective dynamics of GFT condensates, where isotropy is obtained from equilateral tetrahedra. The brown line represents $(V'/3V)^2$, and anisotropic part is shown by the blue or green dot-dashed line. The effective matter term $4\pi G_{eff}/3$ is illustrated by the red dashed line.}
\end{figure}

\begin{figure}[!h]
	\centering 
	\includegraphics[width=10cm]{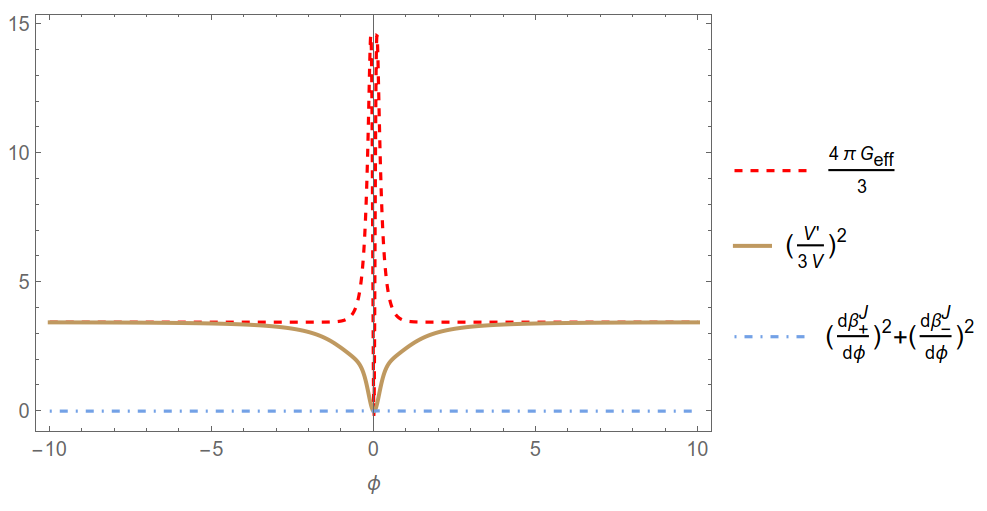}
	\includegraphics[width=9cm]{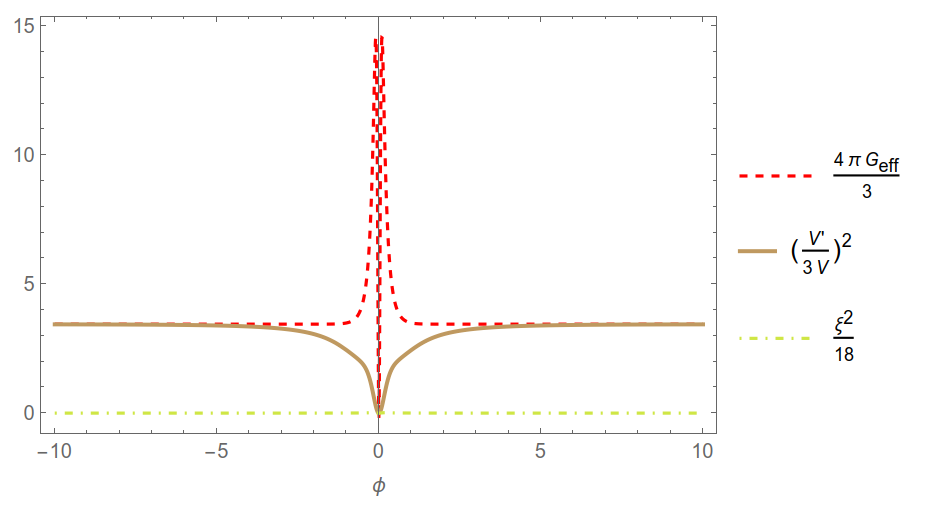}
	\caption{\label{dyn2} Effective dynamics of GFT condensates, where isotropy is obtained from tri-orthogonal tetrahedra. The brown line represents $(V'/3V)^2$, and anisotropic part is shown by the blue or green dot-dashed line. The effective matter term $4\pi G_{eff}/3$ is illustrated by the red dashed line.}
\end{figure}

It is interesting to extract information during the period where anisotropy is relevant after bounce before $\phi\simeq 10$. According to Fig.\ref{dyn} and \ref{dyn2}, one finds that in the regime $5\lesssim\abs{\phi}\lesssim10$, where anisotropy is small but ineligible, an effective Friedmann equation is satisfied, which is a special case of (\ref{friedmann}) with vanishing shear.
Only very small contributions (although non-zero) from anisotropic building blocks can be found, which agrees with the result in \cite{Andrea_2022}.\\

\begin{figure}[!h]
	\centering 
	\includegraphics[width=10cm]{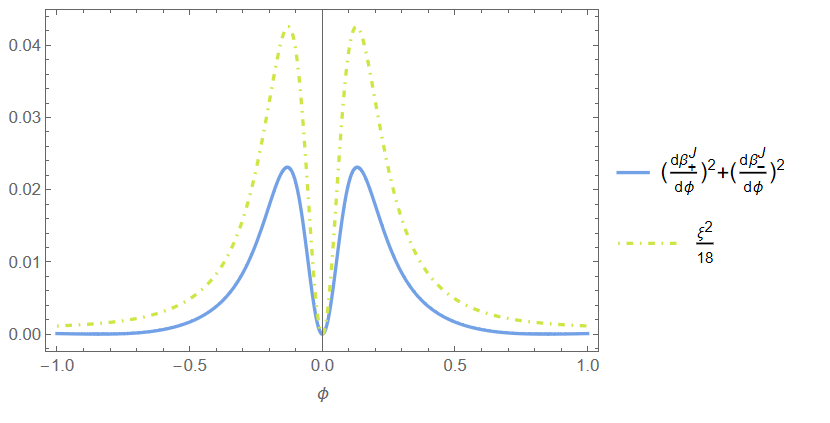}
	\includegraphics[width=10cm]{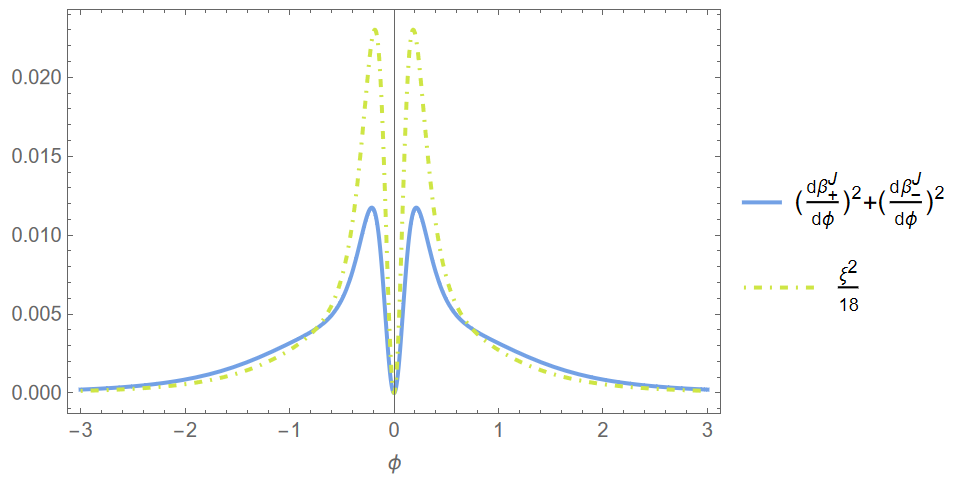}
	\caption{\label{anicompare} Comparison between anisotropic contributions from eqn.(\ref{eff}) and eqn.(\ref{eff2}). The picture above corresponds to the condensates described in Fig. \ref{dyn}, and the one below illustrates the case in Fig. \ref{dyn2}. Though they are equivalent at classical level, the definitions in GFT cosmology makes them slightly different, and the anisotropic part of eqn.(\ref{eff}) \textcolor{red}{is} larger than that of eqn.(\ref{eff2}) near the bounce. Despite the  small difference, the qualitative behaviour of them is the same. To be more precise, Fig. \ref{derivatives} below will show that the derivatives of anisotropic contributions in these two diagrams becomes almost zero even when these contributions are non-vanishing. This can be evidence that the clasical Bianchi I dynamics is approximately satisfied before a transition to isotropic dynamics.}
\end{figure}

\begin{figure}[!h]
	\centering 
	\includegraphics[width=10cm]{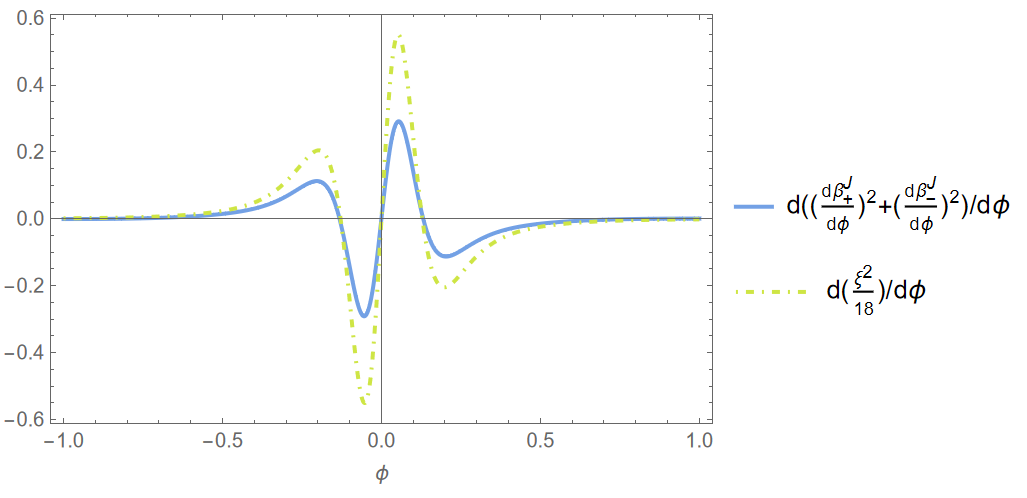}
	\includegraphics[width=10cm]{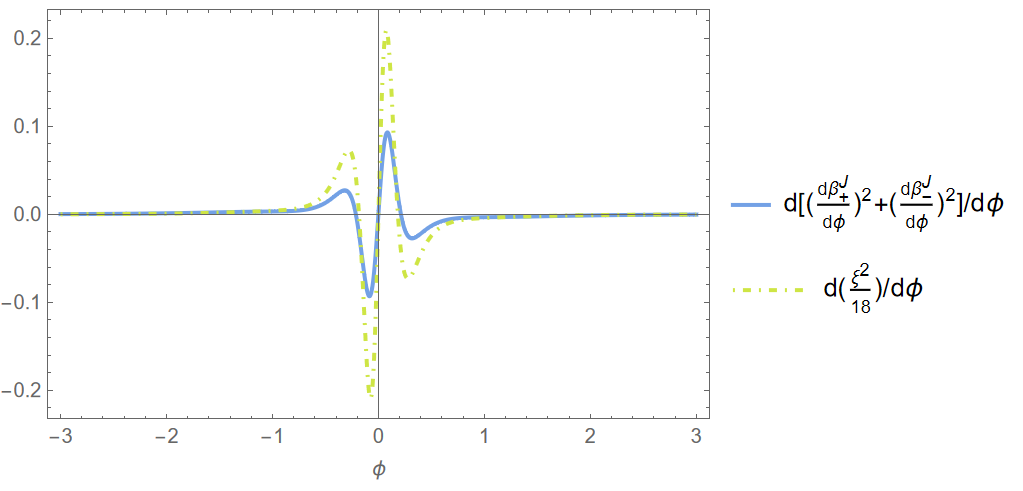}
	\caption{\label{derivatives} Derivatives of anisotropic contributions in eqn.(\ref{eff}) and eqn.(\ref{eff2}) with respect to clock time $\phi$. The picture above corresponds to the condensates described in Fig. \ref{dyn}, where the derivative becomes negligible ($\ll 1$) around $\phi=\pm0.5$. The one below shows the derivatives of anisotropic part in Fig. \ref{dyn2}. Similarly, the derivative becomes negligible around $\phi=\pm0.5$. According to Fig. \ref{anicompare}, these contributions are non-zero when their time derivatives are almost vanishing. These two graphs implies that there is a period where the Bianchi I dynamics is satisfied before the condensates becomes isotropic. }
\end{figure}

Finally let us take a closer look at the period where the condensates near the bounce $-5<\phi<5$ to verify if the small contribution from anisotropy can satisfy the Bianchi I dynamics around $|\phi|\simeq 5$. From Fig. \ref{beta} and Fig. \ref{beta2}, one finds that there is a period where $\beta_{\pm}^J$ varies linearly in time approximately at $1\lesssim|\phi|\lesssim 5$.
We can also see that the slopes are quantitatively very small, so their contribution seems to be almost zero in Fig. \ref{dyn} and Fig.\ref{dyn}. However, according to Fig. \ref{beta} and Fig. \ref{beta2}, it is reasonable to think of the anisotropic contributions to effective dynamics during this period as small but non-trivial.
As Fig. \ref{anicompare} illustrates, the anisotropic part of the dynamics in Fig. \ref{dyn} and Fig. \ref{dyn2} are actually non-zero. Two quantities, $(d\beta_+^J/d\phi)^2+(d\beta_-^J/d\phi)^2$ and $\xi^2/18$, which are equivalent in classical model, share the same qualitative behaviour, with quantitative value slightly different from each other as $|\phi|$ grows. From Fig. \ref{anicompare}, we note that the lower graph (that illustrates the dynamics of tri-othogonal tetrahedra) shows a better agreement betweeen dynamics based on eqn.(\ref{eff}) and eqn.\eqref{eff2}. Meanwhile, the time derivatives of $(d\beta_+/d\phi)^2+(d\beta_-/d\phi)^2$ and $\xi^2/18$ indicates that anisotropic part of GFT effective dynamics is a non-zero constant (though very small) before $|\phi|\simeq 5$, as plotted in Fig. \ref{derivatives}. Therefore, despite the small contribution, a Bianchi I dynamics is approximately satisfied before the condensate becomes dominated by isotropic building blocks.

\subsection{Discussion}
From the calculation above, one finds that the Friedmann equation of a flat FLRW universe is recovered as the number of building blocks increases. However, even before the isotropic background dominates, little correction is brought by anisotropic perturbations. \\

The results need further investigations to be explained in full. However, we can make a few comments. First of all, the collective behaviour expected to bring about a close match with the GR dynamics is valid only when we have a large number of GFT building blocks, but this condition may not be satisfied at early cosmological stage (corresponding to small volumes). Therefore, it is possible that the number of tetrahedra is not large enough to reproduce Bianchi I dynamics (thus GR) effectively.\\ 

We have to recall also that we have used a definition of ``anisotropy'' as a quantum observable according to (\ref{ani1}) and (\ref{ani2}), where an average over all tetrahedra is applied in the first step. This average makes the quantity almost a constant, producing a relative suppression with respect to $V'/V$ of its derivative with respect to $\phi$. The same result is found in \cite{Andrea_2022}. 
One could argue that such an average, thus such suppression, is inevitable because $\beta_{\pm}$ and $k_i$ are intensive quantities. This means that we have to divide them by number operator in definition, while the numerator is proportional to the number operator. Consequently, these quantities are naturally constant, and this is a result heavily depending on the limitation of the construction in this article.
Therefore, this suggest that one may need to find a better way to characterise an anisotropic GFT state. More work to elucidate these issues is anyway needed.\\ 


We have also found that the dominance of isotropic background is a generic result of the mean field GFT dynamics. From eqn.(\ref{soln}), which is a general solution of Schwinger-Dyson equation (\ref{eom}) regardless of the kinetic kernel, one knows the sign of $B_j/A_j$ is important. 
If $B_j/A_j<0$, then the solutions are oscillating. For positive $B_j/A_j$, when $\phi$ is large, $\rho_j\propto\exp{\sqrt{B_j/A_j}\phi}$ approximately. In order for small spins to contribute the most,  $B_j/A_j$ must decrease as spins increase. For example, in this paper we have considered $B_j/A_j=\jmath=-M^2-\eta\sum_i j_i(j_i+1)>0$. We choose $-M^2>0$ and $\eta>0$, so $\jmath$ becomes smaller for larger spins. If one uses other kinetic kernels, it is still immediate to see which modes will dominate by simply checking if $B_j/A_j$ is entirely a non-increasing function of $j$ and it should be positive at least for $j=1/2$.

One last comment is about the isotropisation of GFT condensates.The effective dynamics of GFT condensates shows a transition from anisotropy to isotropy, as they captures the basic features of classical dynamics. However, there are subtle differences between GFT effective dynamics and its classical correspondence. Eqn.\eqref{f1} shows how anisotropy modifies a Friedmann equation via a shear term $\Sigma$. Generally speaking, the isotrpisation of a Bianchi I universe with massless scalar fields is realised by taking $\Sigma^2/a^6\to 0$, where $\Sigma^2$ is fixed and $a^6$ grows in time. In contrast, GFT condensates dynamics becomes isotropic by making $\Sigma^2$ itself increasingly smaller than $a^6$ as $a$ becomes larger. The isotropisation of GFT condensates in this paper is thus stronger than a classical Bianchi I universe. 

\section{Conclusion} 
In this paper, we used GFT quantum gravity condensates to build a homogeneous anisotropic model which reproduces FLRW universe at late time. In this model, a bouncing universe has anisotropic ``perturbations'' at early stage, but it becomes isotropic quickly. 
Compared with previous research on anisotropic GFT cosmology \cite{Cesare_2017, Andrea_2022}, with which we share several qualitative aspects such as the definition of anisotropic condensate states  \cite{Cesare_2017}, and the perturbative treatment of anisotropies \cite{Andrea_2022}, our model has the following crucial improvements:
\begin{itemize}
	\item[i.] The coherent peaked state is used, to extract the relational cosmological dynamics, and to control the quantum fluctuations of observables.
	\item[ii.] All parameters in the GFT wave-function are fixed such that the low-spin phase is dominant at large $\phi$.
	\item[iii.] Misner's variables $\beta_{\pm}$ are both non-vanishing in this paper, and all possible quantities measuring the anisotropy are considered.
	\item[iv.] A more general result is obtained in this paper by summing over all possible modes with constraints on spins.
\end{itemize}


In addition to the Bianchi I dynamics considered in this paper, and partially reproduced by the emergent cosmological dynamics (albeit with all the limitations we have discussed in the previous section), it would be nice to perform a similar analysis for the Bianchi IX model, where the spatial curvature is non-zero. Reproducing its mixmaster dynamics and chaotic behaviour near the singularity from the perspective of the full quantum gravity formalism would be very exciting. 
The main challenge to move in this direction is the definition and inclusion of spatial curvature in the formalism. \\


A main limitation of our analysis was that we worked in the regime of subdominant GFT interactions. This is reasonable for very small volumes (GFT density), but it may still be the case that GFT interactions affect even the early universe dynamics, in particular in its anisotropic aspects. Moreover, GFT interactions are also expected to contribute curvature terms, due to the connectivity between tetrahedra that encode.
In this paper, the isotropic Friedmann dynamics can be well reproduces at late time, but there is no restriction ruling out the curvature terms at early stage, especially when one approaches to the initial singularity. Therefore, though we successfully show the reproduction of Friedmann equation from anisotropic GFT condensates, inclusion of interactions is still called for in the future.
However, the computational challenges involved in including them in the picture are not few and it is beyond the scope of this article. \\

In summary, we have taken a further step toward more realistic cosmology from GFT quantum gravity. Admittedly, many more steps need to be taken to construct solid GFT phenomenology in cosmological content in the future . 



\acknowledgments

The authors would like to thank Andrea Calcinari, Steffen Gielen, and Andreas Phithis for the helpful discussion at the initial stage of this project, and Pi Shi for insights on anisotropic universes. The authors acknowledge financial support from ``Overseas Visiting Fellow Program of Shanghai University''. DO acknowledges financial support from the Deutsche Forschung Gemein-schaft (DFG) via the grants OR 432/3-1 and OR 432/4-1, from the J. Templeton Foundation via the grant 62421, and from the ATRAE programme of the Spanish Government through the grant PR28/23 ATR2023-145735.


\bibliographystyle{JHEP}
\bibliography{GFT}
\end{document}